\documentclass[journal]{IEEEtran}
\usepackage{amsmath,amsfonts}
\usepackage{algorithmic}
\usepackage{algorithm}
\usepackage{array}
\usepackage[caption=false,font=normalsize,labelfont=sf,textfont=sf]{subfig}
\usepackage{textcomp}
\usepackage{stfloats}
\usepackage{url}
\usepackage{verbatim}
\usepackage{makecell}
\usepackage{graphicx}
\usepackage{multirow}
\usepackage{multicol}
\usepackage{booktabs}
\usepackage{xcolor}
\usepackage{hyperref}
\usepackage{cite}
\usepackage{comment}
\usepackage{enumitem}
\usepackage{amssymb}
\usepackage{bm}
\usepackage{subcaption} 

\usepackage{tikz}
\usetikzlibrary{mindmap}
\usepackage{smartdiagram}
\usesmartdiagramlibrary{additions}
\usepackage{forest}
\usetikzlibrary{shadows}
\usepackage{relsize}
\usepackage{color}
\definecolor{lightcoral}{rgb}{0.94, 0.5, 0.5}
\definecolor{lightgreen}{rgb}{0.56, 0.93, 0.56}
\definecolor{lightyellow}{rgb}{0.94, 0.84, 0.6}
\definecolor{brightlavender}{rgb}{0.75, 0.58, 0.89}

\definecolor{skyblue}{rgb}{0.53, 0.81, 0.92}
\definecolor{peachpuff}{rgb}{1.0, 0.85, 0.73}
\definecolor{goldenrod}{rgb}{0.85, 0.65, 0.13}
\definecolor{orchid}{rgb}{0.85, 0.44, 0.84}
\definecolor{salmon}{rgb}{0.8, 0.5, 0.45}
\definecolor{turquoise}{rgb}{0.25, 0.88, 0.82}
\definecolor{plum}{rgb}{0.87, 0.63, 0.87}
\definecolor{khaki}{rgb}{0.94, 0.9, 0.55}
\definecolor{slateblue}{rgb}{0.42, 0.35, 0.8}
\definecolor{forestgreen}{rgb}{0.13, 0.55, 0.13}
\definecolor{midnightblue}{rgb}{0.1, 0.1, 0.44}
\definecolor{lightsteelblue}{rgb}{0.69, 0.77, 0.87}

\definecolor{limegreen}{rgb}{0.2, 0.8, 0.2}
\definecolor{palegreen}{rgb}{0.6, 0.98, 0.6}
\definecolor{springgreen}{rgb}{0.0, 1.0, 0.5}
\definecolor{mediumseagreen}{rgb}{0.24, 0.7, 0.44}
\definecolor{seagreen}{rgb}{0.18, 0.55, 0.34}
\definecolor{yellowgreen}{rgb}{0.6, 0.8, 0.2}
\definecolor{olivedrab}{rgb}{0.42, 0.56, 0.14}
\definecolor{darkseagreen}{rgb}{0.56, 0.74, 0.56}
\definecolor{lightseagreen}{rgb}{0.13, 0.7, 0.67}
\definecolor{forestgreen}{rgb}{0.13, 0.55, 0.13}
\definecolor{darkolivegreen}{rgb}{0.33, 0.42, 0.18}
\definecolor{greenyellow}{rgb}{0.68, 1.0, 0.18}
\definecolor{chartreuse}{rgb}{0.5, 1.0, 0.0}
\definecolor{mintgreen}{rgb}{0.6, 1.0, 0.6}

\begin{document}

\title{Recent Advances in Discrete Speech Tokens: A Review}

\author{Yiwei Guo,~\IEEEmembership{Student Member,~IEEE}, Zhihan Li, Hankun Wang,  Bohan Li, Chongtian Shao, Hanglei Zhang, Chenpeng Du, Xie Chen, Shujie Liu, Kai Yu,~\IEEEmembership{Fellow,~IEEE}

\thanks{Corresponding Author: Kai Yu. Email: kai.yu@sjtu.edu.cn}
\thanks{Yiwei Guo, Zhihan Li, Bohan Li, Chongtian Shao, Hanglei Zhang, Hankun Wang, Chenpeng Du, Xie Chen and Kai Yu are with the MoE Key Lab of Artificial Intelligence, Jiangsu Key Lab of Language Computing; X-LANCE Lab, Department of Computer Science and Engineering, Shanghai Jiao Tong University,
Shanghai, China. Email: yiwei.guo@sjtu.edu.cn}
\thanks{Shujie Liu is with Microsoft Research Asia (MSRA), Beijing 100080, China. 
}
}



\maketitle

\begin{abstract}
The rapid advancement of speech generation technologies in the era of large language models (LLMs) has established discrete speech tokens as a foundational paradigm for speech representation. These tokens, characterized by their discrete, compact, and concise nature, are not only advantageous for efficient transmission and storage, but also inherently compatible with the language modeling framework, enabling seamless integration of speech into text-dominated LLM architectures. Current research categorizes discrete speech tokens into two principal classes: \textit{acoustic} tokens and \textit{semantic} tokens, each of which has evolved into a rich research domain characterized by unique design philosophies and methodological approaches. This survey systematically synthesizes the existing taxonomy and recent innovations in discrete speech tokenization, conducts a critical examination of the strengths and limitations of each paradigm, and presents systematic experimental comparisons across token types. Furthermore, we identify persistent challenges in the field and propose potential research directions, aiming to offer actionable insights to inspire future advancements in the development and application of discrete speech tokens.
\end{abstract}


\IEEEpubidadjcol

\section{Introduction}



\IEEEPARstart{T}{he} rapid advancement of large language models (LLMs) in natural language processing has revolutionized speech generation tasks~\cite{cui2024recent,ji2024wavchat}, with speech being tokenized and modeled using decoder-only Transformers~\cite{transformer}. 
Efforts starting from GSLM~\cite{lakhotia2021generative} and AudioLM~\cite{borsos2023audiolm} aim to develop text-free spoken LLMs, akin to how current LLM-powered chatbots enable text-based interactions. 
Other works, including VALL-E~\cite{valle} and VioLA~\cite{wang2024viola}, extend this approach to conditional speech generation tasks, such as zero-shot text-to-speech and speech translation.
However, this paradigm requires data to be tokenized, as LLMs typically process discrete data only. 
Textual tokens naturally meet this requirement because they are designed as discrete units separated by clear boundaries, whereas raw speech signals are continuous and boundary-less. 
Therefore, a necessary step before applying speech data to LLM is the \textbf{tokenization of speech}, whose goal is:

\begin{center}
    \textit{To transform long speech waveforms into compact sequences of discrete tokens compatible with textual representations, particularly for language modeling tasks involving speech.}
\end{center}

\begin{figure}
    \centering
    \includegraphics[width=0.99\linewidth]{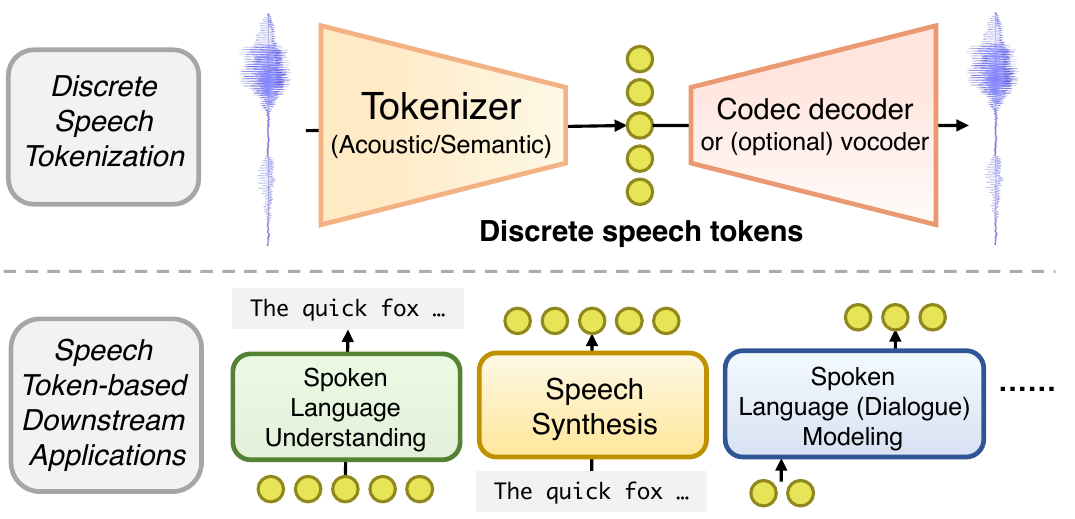}
    \caption{Diagram of discrete speech tokenization process and speech token-based downstream applications.}
    \label{fig:diagram}
    \vspace{-0.2in}
\end{figure}

As a result, significant efforts have been directed towards developing efficient and powerful speech tokenization methods. Generally, these methods are based on two distinct principles, giving rise to two types of speech tokens: \textit{acoustic tokens} and \textit{semantic tokens}.
Acoustic tokens are derived from neural codecs designed to encode speech at a low bitrate while preserving as much information as possible~\cite{wu2024towards,kim2024neural,anees2024speech,du2025codecfake}. In contrast, semantic tokens originate from speech self-supervised learning (SSL)~\cite{mohamed2022self}, which aims to learn a more phonetic or semantic representation space, making it easier for speech recognition.
These two nearly independent lines of research magically intersect in the context of language modeling for speech.
Now, there are also efforts that try to design a speech tokenizer that accomplishes the two objectives simultaneously~\cite{zhang2024speechtokenizer,kyutai2024moshi}.
Consequently, speech tokenization has become a core problem of speech processing under the new paradigm, with versatile downstream applications, as shown in Fig.\ref{fig:diagram}.

Prior to discrete tokens, continuous speech representations from autoencoders and self-supervised learning have been extensively explored~\cite{mohamed2022self}.
Comparatively, continuous representations generally provide higher fidelity in reconstruction or stronger performance on understanding tasks, but lack the compactness, symbolic abstraction, and compatibility with language model-based generation that discrete tokens easily afford. 
In other words, they highlight different tradeoffs in representation learning.


Despite the rapid development and numerous recent works, a comprehensive taxonomy of methodologies in discrete speech tokens has not been clearly constructed. 
Existing reviews~\cite{mohamed2022self,anees2024speech,wu2024towards,cui2024recent,kim2024neural,ji2024wavchat} in this field overlook the diverse categories and methodologies in both acoustic and semantic tokens.
For example, \cite{cui2024recent,ji2024wavchat} focus primarily on methods in spoken language modeling, providing only brief descriptions of some speech tokens used in existing models.
The taxonomy of neural audio codecs has been summarized in \cite{wu2024towards,du2025codecfake}, but the realm of semantic tokens is still overlooked.
In this review, we provide a comprehensive overview of the concepts, methods, and characteristics of various types of discrete speech tokens, with their applications in spoken language understanding, speech generation, and spoken dialogue models.
We hope that through this review, the community can have a clear understanding of the current development and key technologies of discrete speech tokens, so as to promote further research in the future.


Our contributions are summarized as follows:
\begin{itemize}
    \item This review is the first to focus specifically on discrete speech tokens with sufficient depth in the LLM era.
    \item We construct a comprehensive taxonomy of current research on discrete speech tokens and meticulously review the motivation, representative approaches, and challenges in each sub-category.
    \item We provide a unified comparison of different types of discrete speech tokens in terms of reconstruction and voice conversion performance, covering both acoustic and semantic tokens.
    \item We summarize the current challenges and potential future directions for discrete speech tokens, including decoupled and variable frame rate tokens.
\end{itemize}

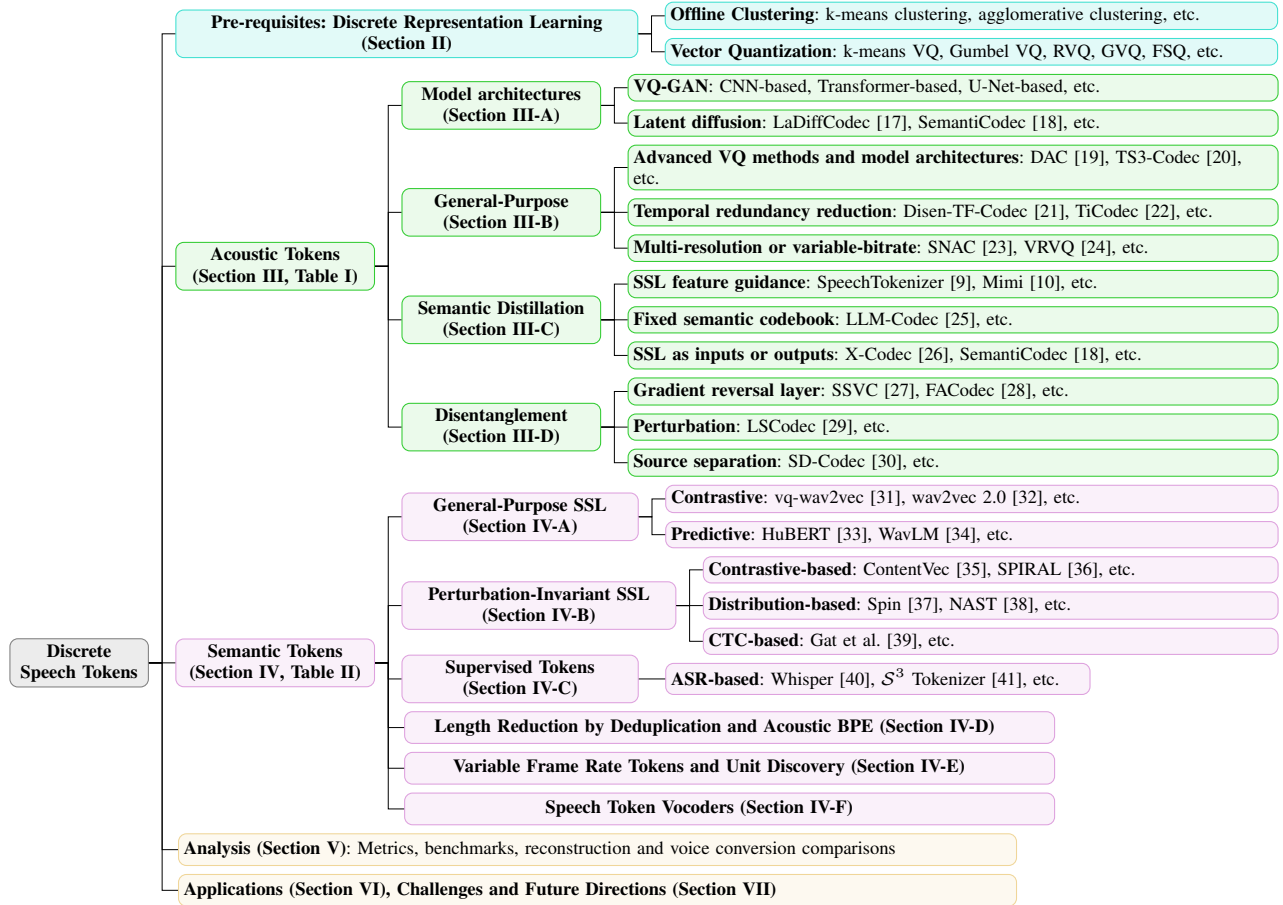
\begin{figure*}[!t]
  \centering
  \tikzset{
          my node/.style={
              draw,
              align=center,
              thin,
              text width=1.2cm, 
              rounded corners=3,
          },
          my leaf/.style={
              draw,
              align=left,
              thin,
              text width=8.5cm, 
              rounded corners=3,
          }
  }
  \forestset{
    every leaf node/.style={
      if n children=0{#1}{}
    },
    every tree node/.style={
      if n children=0{minimum width=1em}{#1}
    },
  }
  \begin{forest}
      nonleaf/.style={font=\bfseries\scriptsize},
       for tree={%
          every leaf node={my leaf, font=\scriptsize},
          every tree node={my node, font=\scriptsize, l sep-=4.5pt, l-=1.pt},
          anchor=west,
          inner sep=2pt,
          l sep=10pt, 
          s sep=3pt, 
          fit=tight,
          grow'=east,
          edge={ultra thin},
          parent anchor=east,
          child anchor=west,
          if n children=0{}{nonleaf}, 
          edge path={
              \noexpand\path [draw, \forestoption{edge}] (!u.parent anchor) -- +(5pt,0) |- (.child anchor)\forestoption{edge label};
          },
          if={isodd(n_children())}{
              for children={
                  if={equal(n,(n_children("!u")+1)/2)}{calign with current}{}
              }
          }{}
      }
      [Discrete Speech Tokens, draw=gray, fill=gray!15, text width=1.5cm, text=black, font=\bfseries, inner ysep=5pt
      [
        Pre-requisites: Discrete Representation Learning \\ (Section \ref{sec:prereq}), color=turquoise, fill=turquoise!15, text width=6cm, text=black
        [{\textbf{Offline Clustering}: k-means clustering, agglomerative clustering, etc.}, color=turquoise, fill=turquoise!15, text width=8cm, text=black]
        [{\textbf{Vector Quantization}: k-means VQ, Gumbel VQ, FSQ, GVQ, RVQ, etc.}, color=turquoise, fill=turquoise!15, text width=8cm, text=black
        ]
      ]
      [{Acoustic Tokens \\ ({Section \ref{sec:acoustic}}, Table \ref{tab:acoustic-metadata})}, color=limegreen, fill=limegreen!10, text width=2.5cm, text=black
            [{Model architectures\\ ({Section \ref{sec:acoustic-arch})}}, 
            color=limegreen, fill=limegreen!10, text width=2.5cm, text=black
            [{\textbf{VQ-GAN}: CNN-based, Transformer-based, U-Net-based, etc.
            }, color=limegreen, fill=limegreen!10, text width=8.6cm, text=black],
            [{\textbf{Diffusion}: LaDiffCodec~\cite{yang2024generative}, SemantiCodec~\cite{liu2024semanticodec}, etc.
            }, color=limegreen, fill=limegreen!10, text width=8.6cm, text=black]
            ],
            [General-Purpose\\ (Section {\ref{sec:acoustic-general}}), color=limegreen, fill=limegreen!10, text width=2.5cm, text=black
            [{
                \textbf{Advanced VQ methods and model architectures}: DAC~\cite{kumar2024high}, TS3-Codec~\cite{wu2024ts3codectransformerbasedsimplestreaming}, etc.
                }, color=limegreen, fill=limegreen!10, text width=8.6cm, text=black],
            [{
                \textbf{Temporal redundancy reduction}: Disen-TF-Codec~\cite{jiang2023disentangled}, TiCodec~\cite{ticodec}, etc.
                }, color=limegreen, fill=limegreen!10, text width=8.6cm, text=black],
            [{
                \textbf{Multi-resolution or variable-bitrate}: SNAC~\cite{Siuzdak_SNAC_Multi-Scale_Neural_2024}, VRVQ~\cite{chae2024variable}, etc.
                }, color=limegreen, fill=limegreen!10, text width=8.6cm, text=black]
            ],
            [Semantic Distillation\\ ({Section \ref{sec:acoustic-distillation}}), color=limegreen, fill=limegreen!10, text width=2.5cm, text=black
              [{ 
                \textbf{Semantic feature guidance}: SpeechTokenizer~\cite{zhang2024speechtokenizer}, Mimi~\cite{kyutai2024moshi}, etc.
              }, color=limegreen, fill=limegreen!10, text width=8.6cm, text=black],
              [{ 
                \textbf{Fixed semantic codebook}: LLM-Codec~\cite{yang2024uniaudio15}, etc.
              }, color=limegreen, fill=limegreen!10, text width=8.6cm, text=black],
              [{ 
                \textbf{Semantic features as inputs or outputs}: X-Codec~\cite{ye2024codec}, SemantiCodec~\cite{liu2024semanticodec}, etc.
              }, color=limegreen, fill=limegreen!10, text width=8.6cm, text=black],
            ],
            [Disentanglement\\ ({Section \ref{sec:acoustic-disen}}), color=limegreen, fill=limegreen!10, text width=2.5cm, text=black
              [{ 
                \textbf{Gradient reversal layer}: SSVC~\cite{SSVC}, FACodec~\cite{facodec}, etc.
              }, color=limegreen, fill=limegreen!10, text width=8.6cm, text=black],
              [{ 
                \textbf{Perturbation}: LSCodec~\cite{guo2024lscodec}, etc.
              }, color=limegreen, fill=limegreen!10, text width=8.6cm, text=black],
              [{ 
                \textbf{Source separation}: SD-Codec~\cite{bie2024learning}, DeCodec~\cite{luo2025decodec}, etc.
              }, color=limegreen, fill=limegreen!10, text width=8.6cm, text=black],
            ],
          ]
    [Semantic Tokens \\ (Section {\ref{sec:semantic}, Table \ref{tab:semantic-metadata}}), color=plum, fill=plum!15, text width=2.5cm, text=black
            [SSL Semantic Tokens\\(Section {\ref{sec:semantic-general}}), color=plum, fill=plum!15, text width=3cm, text=black
            [{\textbf{Contrastive models}: vq-wav2vec~\cite{vq-wav2vec}, wav2vec 2.0~\cite{baevski2020wav2vec}, etc.}, color=plum, fill=plum!15, text width=8.1cm, text=black],
            [{\textbf{Predictive models}: HuBERT~\cite{hsu2021hubert}, WavLM~\cite{chen2022wavlm}, etc.}, color=plum, fill=plum!15, text width=8.1cm, text=black],
            [{\textbf{Perturbation-invariant SSL models}: ContentVec~\cite{qian2022contentvec}, NAST~\cite{messica2024nast}, etc.}, color=plum, fill=plum!15, text width=8.1cm, text=black]
          ],
          [{\textbf{Supervised Semantic Tokens (Section {\ref{sec:semantic-supervised}})}: Whisper~\cite{whisper}, $\mathcal S^3$ Tokenizer~\cite{du2024cosyvoice}, etc.}, color=plum, fill=plum!15, text width=8.7cm, text=black
          ],
          [\textbf{Speech Token Vocoders (Section {\ref{sec:vocoder}})}, color=plum, fill=plum!15, text width=8.7cm, text=black, align=center
          ],
      ],
      [Length Reduction and Variable-Rate Tokenization \\ (Section {\ref{sec:length-and-vfr}}), color=salmon, fill=salmon!15, text width=5.5cm, text=black
        [\textbf{~~~~Length Reduction by Deduplication and Acoustic BPE ({Section \ref{sec:dedup-bpe}})}, color=salmon, fill=salmon!15, text=black, align=center, text width=8.5cm
          ],
          [\textbf{~~~~~~~Variable Frame Rate Tokens and Unit Discovery (Section {\ref{sec:variable-rate}})}, color=salmon, fill=salmon!15, text width=8.5cm, text=black, align=center
          ],
      ]
      [{\textbf{Analysis (Section \ref{sec:analysis})}: Metrics, benchmarks, experimental comparisons (reconstruction, voice conversion, downstream semantic modeling)}, color=lightyellow, fill=lightyellow!15, text width=14.5cm, text=black, align=center
       ],
       [\textbf{Applications (Section \ref{sec:application}), Challenges and Future Directions (Section \ref{sec:challenge})}, color=lightyellow, fill=lightyellow!15, text width=14.5cm, text=black, align=center
       ]
       ]
      \end{forest}
  \caption{Structure of this review. 
  After a brief introduction to the preliminary knowledge, we will taxonomize acoustic tokens and semantic tokens, followed by cross-cutting methods such as length reduction and variable-rate tokenization. 
  Later sections cover experimental analysis, applications, and future directions. Each branch corresponds to a subsection in the paper.}
  \vspace{-0.2in}
  \label{fig: Taxonomy}
  \end{figure*}

The structure of this review is shown in Fig.\ref{fig: Taxonomy}.
Following \cite{borsos2023audiolm,kharitonov2023speak,yang2024towards}, we classify discrete speech tokens into acoustic and semantic tokens based on their principles.
We will characterize the two types of tokens both by conceptual descriptions and unified experimental comparisons.

\section{Pre-requisites: Discrete Representation Learning}
\label{sec:prereq}

Discrete speech tokens are obtained through the quantization of continuous representations, which is usually achieved by offline clustering or online vector quantization algorithms.
This section provides a concise overview of the existing quantization methods commonly used in discrete speech tokens.

Denote $\bm x\in \mathbb R^d$ as a vector in the $d$-dimensional continuous space. A quantization process $q$ transforms $\bm x$ into a discrete \textit{token} in a finite set, i.e. $q(\bm x): \mathbb R^d\to\{1,2,...,V\}$ where $V$ is the \textit{vocabulary size}.
The output tokens are sometimes referred to as \textit{indexes} in the finite $V$-cardinal set.
The function $q$ is usually associated with a \textit{codebook} $\mathcal C=\{\bm c_1,\bm c_2,...,\bm c_V\}$ where every \textit{code-vector} $\bm c_i\in\mathbb R^d$ corresponds to the $i$-th token. 
The code-vectors are representations of tokens in the original $d$-dimensional space.
As $V$ elements can be encoded using $\lceil \log_2 V\rceil$ raw bits\footnote{We denote by $\lceil z\rceil$ and $\lfloor z\rfloor$ the ceiling and floor of scalar $z$, i.e., 
$\lceil z\rceil = \min\{n\in\mathbb{Z}:n\ge z\}$ and 
$\lfloor z\rfloor = \max\{n\in\mathbb{Z}:n\le z\}$.
},
 quantization often compresses the cost for data storage and transmission to a great extent.

\vspace{-0.15in}
\subsection{Offline Clustering}
\vspace{-0.05in}
Clustering is a simple approach for quantization. 
Given a dataset $X=\{\bm x_1,\bm x_2,...\bm x_N\}$, a clustering algorithm aims to assign each sample $\bm x_i$ to a group such that some cost is minimized.
The most frequently used clustering method for discrete speech tokens is k-means clustering~\cite{IKOTUN2023178}. For example, k-means is applied on HuBERT~\cite{hsu2021hubert} features in GSLM~\cite{lakhotia2021generative}.
K-means is a clustering algorithm based on Euclidean distances.
Its training process iteratively assigns each data sample to the nearest centroid, and moves cluster centroids till convergence, with a pre-defined number of clusters.
After training, the centroids form the codebook, and new data can be quantized to the index of the nearest centroid in this Voronoi partition.
In practice, centroids are usually initialized with the k-means++ algorithm~\cite{kmeans++} for better convergence.

Hierarchical agglomerative clustering has also been used in discrete speech tokens, which iteratively merges the closest clusters.
It is usually applied after k-means to reduce the number of clusters~\cite{cho2024sd,baade2024syllablelm}.
Other clustering algorithms are less explored in the context of discrete speech tokens.

\vspace{-0.12in}
\subsection{Vector Quantization}
\vspace{-0.04in}

\IEEEpubidadjcol

Clustering is often an isolate process, thus cannot be optimized together with other neural network modules.
Instead, vector quantization (VQ)~\cite{gray1984vector} enables a learnable network module that allows gradients to pass through when producing discrete representations.
Autoencoders with a VQ module is termed VQ-VAE~\cite{VQVAE}.
There are multiple VQ methods:

\subsubsection{K-means VQ}
Like k-means clustering, k-means VQ method finds the code-vector closest to the input, i.e. 
\begin{equation}
    q(\bm x)=\underset{i\in \{1,2,...,V\}}{\arg\min} \|\bm x-\bm c_i\|^2.
\end{equation}
Then, code-vector $\bm c_k\triangleq\bm c_{q(\mathbf x)}$ is fed to subsequent networks.
As the $\min$ operation is not differentiable, straight-through estimators (STEs)~\cite{bengio2013estimating} are usually applied to graft gradients, i.e. $\operatorname{STE}(\bm c_k,\bm x)=\bm x+\operatorname{sg}(\bm c_k-\bm x)$ where $\operatorname{sg(\cdot)}$ stops tracking gradients.
In this way, the input value to subsequent networks is still $\bm c_k$, but gradients are grafted to $\bm x$ in back propagation.

Auxiliary loss functions are often used together with k-means VQ~\cite{VQVAE}: commitment loss $\mathcal L_{\text{cmt}}=\|\operatorname{sg}(\bm c_k)-\bm x\|^2$ and codebook loss $\mathcal L_{\text{code}}=\|\operatorname{sg}(\bm x)-\bm c_k\|^2$.
The commitment loss pushes the continuous input $\bm x$ towards the closest codebook entry, while the codebook loss does the opposite and updates the code-vector $\bm c_k$.
The two loss terms are weighted by different factors to put different optimization strengths on $\bm x$ and $\bm c_k$, as pushing $\bm c_k$ towards $\bm x$ is an easier task.
It is also common to replace $\mathcal L_{\text{code}}$ with exponential moving average (EMA) to update the codebook instead~\cite{razavi2019generating}, which does not rely on explicit loss functions.

VQ in high-dimensional spaces is known to suffer from codebook collapse,  where the codebook usage is highly imbalanced~\cite{lancucki2020robust,dhariwal2020jukebox}.
To improve codebook utilization, random replacement (as known as \textit{codebook expiration}) can be applied~\cite{dhariwal2020jukebox} on code-vectors that have remained inactive for a long time. 
Other popular solutions include additional auxiliary constraints such as entropy penalty~\cite{chang2022maskgit,yu2024language}, factorized codebook lookup in low-dimensional space~\cite{yu2022vectorquantized}, and reparameterizing code-vectors through a learnable linear projection~\cite{zhu2024addressing}. 

\subsubsection{Gumbel VQ}
Instead of quantizing by Euclidean distance, another choice is by probability. 
Gumbel VQ~\cite{jang2017categorical} uses Gumbel-Softmax as a proxy distribution for traditional Softmax to allow differentiable sampling.
Given input $\bm x$ and a codebook of size $V$, a transform $h(\cdot)$ is applied on $\bm x$ into $V$ logits: $\bm l=h(\bm x)\in \mathbb R^V$.
In inference, quantization is performed by choosing the index with the largest logit, i.e. $q(\bm x)=\arg\max_i \left\{\bm l^{(i)}\right\}$.
In training, samples are drawn from the categorical distribution implied by $\bm l$ for the subsequent neural networks.
To achieve efficient sampling and let gradients pass through, Gumbel trick is used:
\begin{align}
    &\bm u\in \mathbb R^V\sim \operatorname{Uniform}(0, 1),\bm v=-\log(-\log(\bm u)) \label{eq:gumbel-noise} \\
    &\bm s=\operatorname{Softmax}((\bm l+\bm v)/\tau) \label{eq:gumbel-softmax}
\end{align}
where Eq.\eqref{eq:gumbel-noise} samples Gumbel noise $\bm v$ element-wise, and Eq.\eqref{eq:gumbel-softmax} calculates Gumbel-Softmax distribution $\bm s$ with a temperature $\tau$.
The forward pass simply uses $j=\arg\max_i \{\bm s^{(i)}\}$ as the sampled index, but the true gradient of Gumbel-Softmax is used in backward pass.
In other words, the gradient on the one-hot distribution corresponding to $j$ is grafted to $\bm s$ as an approximate.
The temperature $\tau$ balances the approximation accuracy and gradient variances.
The transform $h(\cdot)$ is usually parameterized as neural networks, or negatively proportional to Euclidean distances~\cite{jiang2023latent}.

After quantization, code-vector $\bm c_k$ with $k=q(\bm x)$ is fed to subsequent networks.
Gumbel VQ does not require additional losses, since code-vectors can be directly learned with gradients and do not need to be pushed towards $\bm x$.

\subsubsection{Finite Scalar Quantization (FSQ)}
As mentioned before, VQ methods based on code-vector assignment usually suffer from codebook collapse. 
Despite many efforts, this remains a crucial challenge.
FSQ~\cite{mentzer2024finite} is an alternative that performs quantization in scalar domain.
FSQ quantizes each dimension of a vector $\bm x$ into $L$ levels.
For the $i$-th dimension $\bm x^{(i)}$, FSQ transforms the values into a limited range and then rounds to integers, i.e. 
\begin{equation}
    q\left(\bm x^{(i)}\right)=\operatorname{round}\left(\lfloor L/2\rfloor\tanh\left(\bm x^{(i)}\right)\right).
\end{equation}
The quantized values for each dimension are thus integers ranging from $-\lfloor L/2 \rfloor$ to $\lfloor L/2 \rfloor$\footnote{Following \cite{mentzer2024finite}, this is the symmetric case for $L$ being odd. When $L$ is even, there is an offset before rounding to obtain asymmetric quantized values.}.
For a $d$-dimensional vector $\bm x$, there are $V=L^d$ possible quantization outcomes.
STE is also applied to pass gradients.
As quantization is simply done via rounding to integers, there are no explicit codebooks associated with the FSQ process.

\subsubsection{Other VQ Tricks}
In many cases, a single VQ module suffers from a highly-limited representation space, thus results in poor performance compared to continuous counterparts. 
There are some widely-used VQ tricks that introduce multiple quantizers to refine the quantized space, as shown in Fig.\ref{fig:gvq-rvq}:

\begin{figure}
    \centering
    \includegraphics[width=0.99\linewidth]{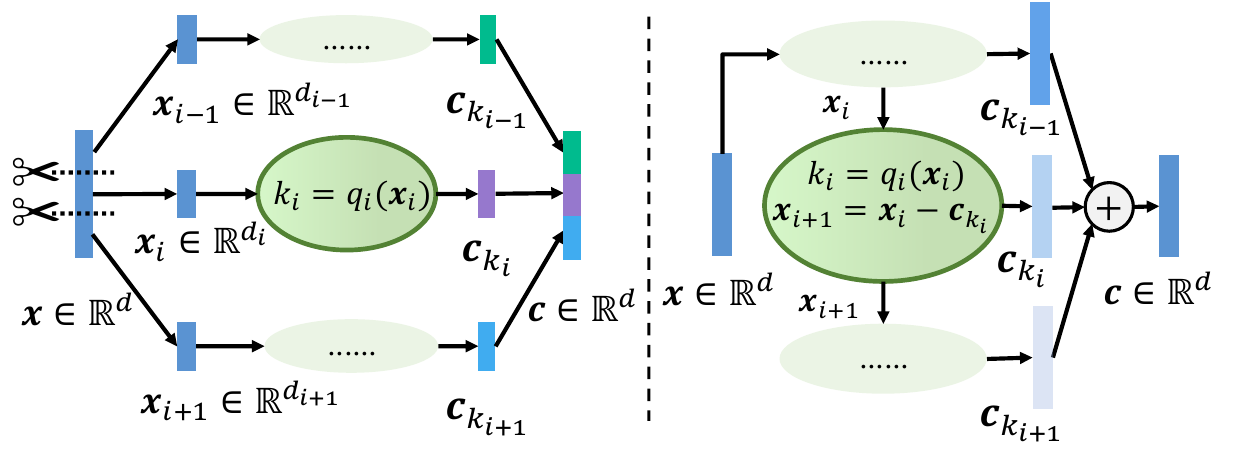}
    \caption{Diagram of GVQ (left) and RVQ (right). GVQ quantizes different partitions of the input vector independently, while RVQ sequentially quantizes the residuals.}
    \label{fig:gvq-rvq}
    \vspace{-0.1in}
\end{figure}

\begin{enumerate}[leftmargin=5mm]
    \item \textit{Grouped VQ (GVQ)}, also known as \textit{product quantization}~\cite{product_quantization}. It partitions the input vector $\bm x$ by dimensions and applies VQ on different groups independently. Different groups can have different or shared codebooks. 
    Each group produces a code-vector of the same dimensionality as its partition. 
    These code-vectors are concatenated across groups to form a final output whose dimensionality equals that of $\bm x$.
    \item \textit{Residual VQ (RVQ)}, also known as \textit{multi-stage quantization}~\cite{multiple-stage-vector-quantization}. It adopts a serial approach that iteratively quantizes the residual of the last quantizer.
    Similar to GVQ, RVQ also has multiple quantizers.
    For the $i$-th quantizer $q_i$ with input $\bm x_i$ and output code-vector $\bm c_{k}$, the residual is defined as $\bm x_{i+1}=\bm x_i-\bm c_{k}$.
    The outputs from all $q_i$ are finally summed as the quantized result of $\bm x$.
    In this way, information in the codebooks is supposed to follow a coarse-to-fine order, and more details in the original $\bm x$ can be preserved than a plain quantizer.
\end{enumerate}
GVQ and RVQ can also be flexibly combined to form GRVQ~\cite{yang2023hifi} that applies RVQ on each GVQ branch for better codebook utilization.
RVQ can also be applied to FSQ~\cite{parker2024scalingtransformerslowbitratehighquality}.
Note that RVQ naturally produces an order of importance in residual layers, while all quantizers in GVQ are equally important.
Such order of importance can also be enforced in GVQ by a ``nested dropout'' trick~\cite{rippel2014learning}.

\subsubsection{Comparisons}
Compared to k-means VQ, both Gumbel VQ and FSQ avoid additional loss terms during training. 
However, Gumbel VQ is sensitive to the temperature parameter $\tau$ in practice~\cite{shah2024improving}. 
FSQ, by contrast, has a simpler design and optimization procedure, and has been reported to achieve better codebook utilization\footnote{Although FSQ does not maintain an explicit codebook, utilization can still be measured over the $V=L^d$ possible outcomes.} under large vocabulary sizes than k-means VQ~\cite{mentzer2024finite}. 

Nevertheless, FSQ also has certain limitations.
In FSQ, as the vocabulary size $V$ follows $V=L^d$, $d$ is usually chosen to be small (like $d=6$ in StableCodec~\cite{parker2024scalingtransformerslowbitratehighquality}).
For VQ, $V$ is not related to the code-vector dimension $d$, which can therefore be set to a wider range of values.
The bottleneck of such low dimensionality might also cause FSQ to underperform in small vocabulary sizes compared to a fully-utilized VQ~\cite{mentzer2024finite}.
Also, the quantization space of FSQ is strictly fixed, whereas VQ methods maintain a learnable codebook. 
This rigidity forces the surrounding network modules, particularly the encoder, to map the data distribution into such low-dimensional structured codes, thereby placing greater demands on model capacity.

Regarding GVQ and RVQ, the inherent ordering in RVQ provides greater flexibility than GVQ in trading off bitrate and performance. 
However, RVQ’s sequential nature not only prevents efficient parallelization in computation, but might also complicate optimization due to the nested STE operations.


\vspace{-0.03in}
\section{Speech Tokenization Methods: Acoustic Tokens}
\label{sec:acoustic}

Acoustic speech tokens are discrete representations derived from \textit{codec models}, primarily designed for speech compression and reconstruction.
The audio codec technology emerged long ago.
Traditional codecs, including 
MP3~\cite{rfc5219}, Opus~\cite{Valin2012DefinitionOT} and EVS~\cite{dietz2015overview}, typically take advantage of signal processing algorithms to improve quality and lower the bitrate.

In the deep learning era, numerous codec models based on neural networks have been developed.
These models typically consist of an encoder that compresses speech signals and a decoder that reconstructs the speech signals, with a quantizer situated between the two.
The quantizer is also parameterized and jointly trained with the whole network in an end-to-end manner. 
The codebook indices produced by the quantizer are referred to as acoustic tokens.
To improve the representation ability of discrete VQ spaces and thus obtain better codec performance, RVQ, GVQ, GRVQ and FSQ tricks are commonly applied in the quantization module.

We list the VQ method, number of quantizers $Q$, frame rate $F$, vocabulary size $V$ for each quantizer, and the resulting bitrate of existing neural acoustic speech tokens in 
Table \ref{tab:acoustic-metadata} in the appendix.

\subsection{Model Architectures}
\label{sec:acoustic-arch}

\begin{figure}
    \centering
    \includegraphics[width=0.99\linewidth]{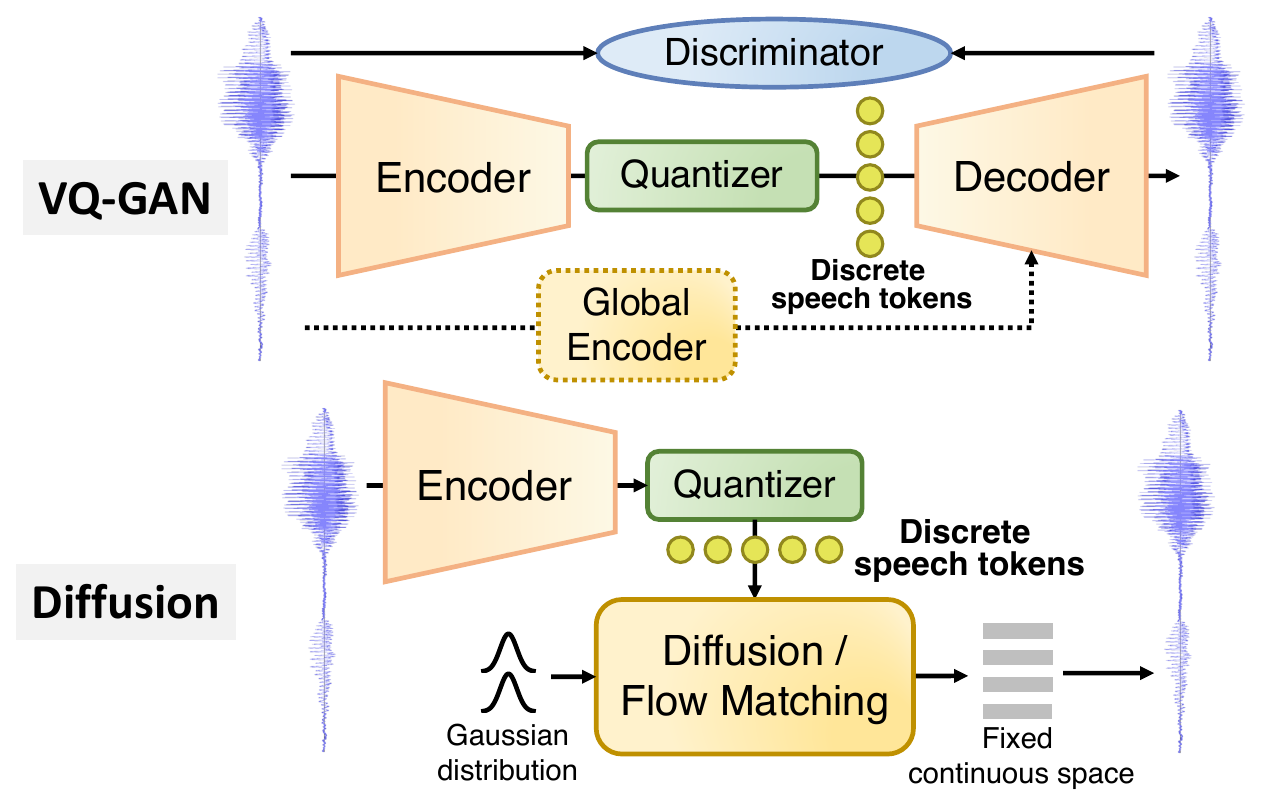}
    \caption{Neural architectures of acoustic tokens.
    Note that inputs and outputs can be waveforms, frequency-domain features or even SSL features depending on purpose and design.}
    \label{fig:acoustic-paradigms}
    \vspace{-0.15in}
\end{figure}

Although acoustic codec models differ from one to one regarding their purposes, most of them share a similar encoder-quantizer-decoder framework.
With audio clip $\bm x$ that can either be time-domain sampling points, frequency-domain features or even other machine learning features, an encoder $f_\theta(\cdot)$ transforms it to $f_\theta(\bm x)$ in a continuous latent vector space. 
The encoder $f_\theta(\cdot)$ will usually perform downsampling to reduce the temporal length of the input signals, especially for waveform inputs.
A VQ module $q_\phi(\cdot)$ discretizes $f_\theta(\bm x)$ into tokens and corresponding codebook vectors $\bm c$.
A decoder $g_\psi(\cdot)$ then uses $\bm c$ to reconstruct $\hat {\bm x}$, and a certain distance metric of $d(\bm x, \hat{\bm x})$ is usually optimized.
There are two major paradigms for designing the encoder, decoder, and quantizers, which can be summarized as diagrams in Fig.\ref{fig:acoustic-paradigms}.

\begin{figure}
    \centering
    \includegraphics[width=0.99\linewidth]{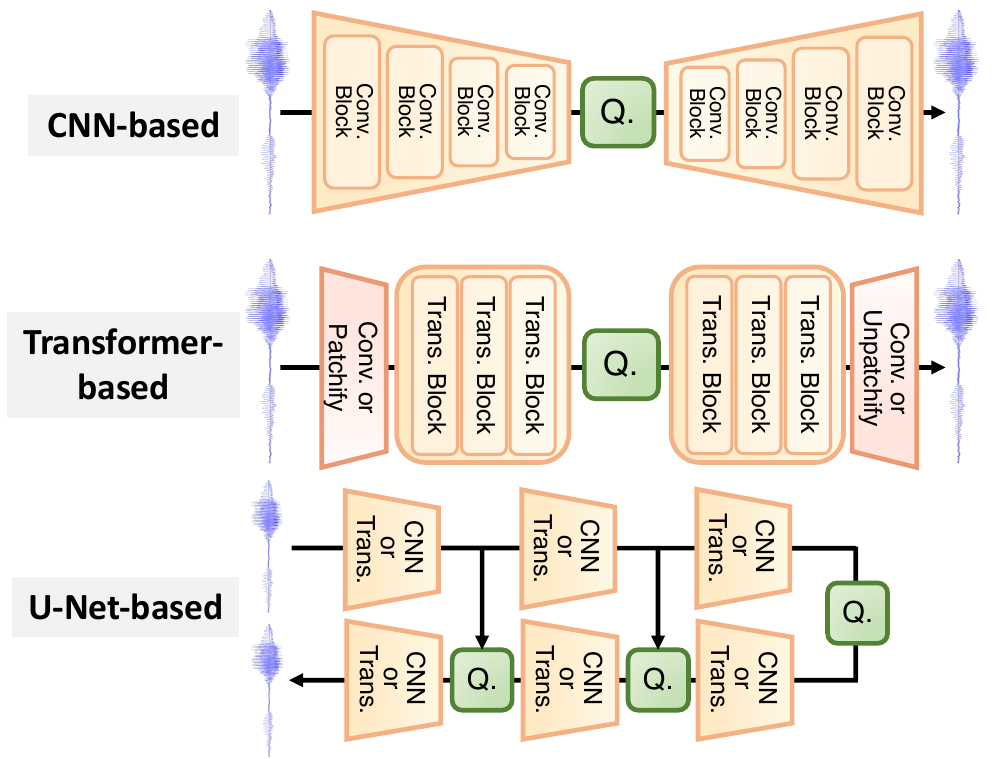}
    \caption{Major generator architectures of VQ-GAN-based acoustic tokens. 
    ``Conv.'', ``Q.'' and ``Trans.'' are short for convolution, quantizer and Transformer, respectively.}
    \label{fig:generator}
    \vspace{-0.2in}
\end{figure}

\subsubsection{VQ-GAN}
VQ-GAN~\cite{esser2021taming} is a very commonly adopted framework of codec models that trains a VQ-VAE with GAN objectives. 
Besides the original reconstruction and VQ objectives in a VQ-VAE, VQ-GAN uses discriminators $d_\xi(\bm x, \hat{\bm x})$ to distinguish real and reconstructed data that adversarially train the generator network composed of $f_\theta,q_\phi$, and $g_\psi$. In acoustic codecs, there are usually multiple discriminators, e.g. multi-resolution and multi-scale STFT discriminators from the neural vocoder researches~\cite{kumar2019melgan,jang21_interspeech}.
The generator architecture of VQ-GAN-based codec models has multiple choices, with the three most representative ones visualized in Fig.\ref{fig:generator}: CNN-based, Transformer-based, and U-Net-based.

The CNN-based generator is the most widely used architecture so far in codec models.
SoundStream~\cite{zeghidour2021soundstream} and EnCodec~\cite{encodec} are two famous early neural codec models that operate in an end-to-end VQ-GAN manner.
They receive time-domain waveforms as inputs and directly reconstruct waveforms.
Their encoder and decoder have a mirrored architecture to perform down and up-samplings.
In SoundStream, the encoder and decoder are purely constructed by convolutional neural networks (CNNs) while EnCodec augments them with an LSTM.
The CNN encoder down-samples the waveform to a high-dimensional embedding sequence, whose frame rate is determined by the sampling rate, CNN kernel sizes and strides at a fixed ratio.
The continuous embeddings are passed through an RVQ quantizer. The resulting quantized vectors are summed and then passed to a CNN decoder to reconstruct the waveform.
The training criteria include reconstruction loss (in the time and frequency domain), adversarial loss, feature matching loss, and quantization losses for RVQ layers.
To allow for a flexible choice of bitrates, structured dropout is adopted where the number of codebooks in the RVQ module can be randomly chosen~\cite{zeghidour2021soundstream}, such that only a portion of quantizers in front are activated during training.
The resulting acoustic tokens can consequently reside in variable bitrates depending on the chosen number of RVQ quantizers.
The inputs and outputs of the codec model can also be frequency-domain features like magnitude and phase spectra for reducing computation burden~\cite{du2024funcodec}.
There, the convolution kernels are typically 2D instead of 1D in the time-domain codecs.

Later, Transformers~\cite{transformer} have been adopted, e.g. in Single-Codec~\cite{singlecodec} and Mimi~\cite{kyutai2024moshi}.
They can be directly applied to frequency-domain inputs and outputs.
When operating on waveform-domain inputs or outputs, a CNN~\cite{kyutai2024moshi} or patchifying~\cite{wu2024ts3codectransformerbasedsimplestreaming,parker2024scalingtransformerslowbitratehighquality} operation is usually added before or after the Transformer blocks.
In Mimi, a shallow Transformer layer is added after the CNN-based encoder, and vice versa in its decoder.
Recently, some propose to use purely Transformer-based backbone and discard the CNN blocks, e.g. TS3-Codec~\cite{wu2024ts3codectransformerbasedsimplestreaming}.
As Transformers demonstrate superior modeling ability and scaling property, these works prove to outperform CNN-based codecs either with less computation~\cite{wu2024ts3codectransformerbasedsimplestreaming} or larger scale~\cite{parker2024scalingtransformerslowbitratehighquality}.
However, to ensure stream-ability, an attention mask should be employed~\cite{kyutai2024moshi}.
The encoder and decoder can also be designed to be different. 
For example, Single-Codec~\cite{singlecodec} uses Conformer~\cite{conformer} encoder and CNN decoder, while LSCodec~\cite{guo2024lscodec} uses the reverse configuration.

In addition, U-Net-based codecs employ multiple quantizers at different layers of the network, rather than relying on a single GVQ or RVQ module for the entire codec. Typical examples in this category include CoFi-Codec~\cite{guo2024speaking} and ESC~\cite{gu2024esc}. In such designs, each sub-encoder or decoder in the U-Net can be implemented with a CNN or Transformer, offering more flexible control over the resolution of each VQ stream (Section~\ref{sec:multi-resolution}).

It is also noteworthy that training a separate vocoder on top of existing acoustic tokens may result in improved audio quality than the original decoded outputs, since reconstructing waveform alone may be simpler than optimizing VQ representation and reconstruction at the same time.
This is exemplarily verified in AudioDec~\cite{audiodec}, MBD~\cite{san2023discrete} and Vocos~\cite{siuzdak2024vocos}.
Therefore, some codec models directly simplify the VQ-GAN training objective back to the original VQ-VAE, where the discrete acoustic tokens are obtained first by a simple reconstruction loss, and a vocoder is trained as an additional stage, like AudioDec~\cite{audiodec} and LSCodec~\cite{guo2024lscodec}.
These works are denoted as ``VQ-VAE+GAN'' in 
Table S-I.


\subsubsection{Diffusion} 
Different from VQ-GAN which uses GAN to generate waveforms or frequency features, some codecs also use denoising diffusion~\cite{ho2020denoising,song2021scorebased} or flow matching models~\cite{lipmanflow} as an alternative.
Since diffusion and flow matching belongs to the same family of generative models, we collectively refer to them as ``diffusion'' throughout this paper.
These diffusion-based codecs use discretized tokens to condition the transformation of standard Gaussian distributions to the distribution of some continuous acoustic features, e.g. spectrogram features, or the latent space of a pretrained speech autoencoder.
The diffusion loss can be propagated back to the encoder and quantizer in this design, like~\cite{liu2024semanticodec}.
The encoding process in such codecs is identical to that of VQ-GAN.
In the decoding process, the decoder runs an iterative sampling process to generate the target acoustic features, which are converted to waveforms by a separate pretrained model.


\subsubsection{Comparisons}
Compared to diffusion, VQ-GAN-based codecs are more intuitive in design, and have been a well-established method.
Within this category, different architectures offer distinct advantages: CNN-based models are usually lightweight and context-invariant because of a limited receptive field.
Transformer-based models are easier to scale and believed to have better compression capacity.
U-Net-based models offer greater flexibility in quantizer resolutions, but the correlation of tokens from different quantizers may be more complex for downstream modeling compared to adjacent quantizers in a single RVQ module.

However, VQ-GAN-based codecs rely on sophisticated discriminators, which is crucial to the performance. 
In contrast, diffusion-based codecs do not need adversarial training, and thus have a simpler training objective.
Inference latency is a major concern of diffusion-based codec models, unless specifically optimized for limited sampling steps.


\vspace{-0.1in}
\subsection{General-Purpose Acoustic Tokens}

\label{sec:acoustic-general}

\subsubsection{Motivation}

In this section, we describe the most common type of neural acoustic tokens (speech codecs) that are designed only with the objective of speech signal reconstruction.
Those acoustic tokens are designed to achieve better objective or perceptual quality at the lowest possible bitrates.

\subsubsection{Approaches}



\paragraph{Advanced VQ methods and model architectures}

Based on SoundStream and EnCodec, more codecs with advanced VQ methods, network structure, or optimization strategies have been researched with depth.
As an example, DAC~\cite{kumar2024high} achieves remarkable reconstruction quality by adding periodic inductive bias, better discriminators, modified loss functions, and a better VQ mechanism from ViT-VQGAN~\cite{yu2022vectorquantized} to improve codebook usage. 
Specifically, it performs L2-normed code lookup in a low-dimensional space (e.g. 8 or 32) instead of a high-dimensional space like 1024.
Other architectural improvements include using frequency-domain inputs~\cite{APCodec,ai24b_interspeech,singlecodec}, variance-constrained residual blocks~\cite{ahn2024hilcodec}, multi-filter bank discriminator~\cite{ahn2024hilcodec}, selective down-sampling back-projection~\cite{zheng2024supercodec}, etc.


Several training tricks are explored, such as not applying VQ with a certain probability and pure adversarial training proposed in Moshi~\cite{kyutai2024moshi}.
Also, the training of neural speech codecs does not need to be end-to-end, i.e. the learning of VQ representations and signal reconstruction can be separated.
\cite{audiodec,du2024apcodec+} adopt a two-stage training process that introduces adversarial losses and an additional vocoder after training only with metric losses, to achieve improved quality.
Additional training criteria on the VQ module are proposed for better VQ utilization.
For example, ERVQ~\cite{zheng2024ervq} introduces a fine-grained code-vector replacement strategy, a codebook balancing loss, and a similarity loss between consecutive RVQ layers.



Other VQ methods besides GVQ or RVQ also exist in speech codecs.
NDVQ~\cite{niu2024ndvq} improves the capacity of RVQ space by changing codebook {vectors} to parameterized Gaussian {distributions}.
FSQ has also been introduced to several speech codecs, like SQ-Codec~\cite{yang24l_interspeech} where scalar rounding is applied to each of its 32-dimensional latent space.
Stable-Codec~\cite{parker2024scalingtransformerslowbitratehighquality} adopts FSQ in a Transformer-based architecture, exhibiting strong scalability to large model sizes up to 950M parameter count.
It also explores a flexible post-training quantization level adjustment technique and residual FSQ strategy.

Note that most acoustic tokens require multiple quantizers, but \textbf{single-codebook} codecs have also been explored.
Single-Codec~\cite{singlecodec} designs an encoder consisting of Conformer and bidirectional LSTM to better compress mel spectrogram inputs.
WavTokenizer~\cite{ji2024wavtokenizer} and BigCodec~\cite{xin2024bigcodec} further explores single-codebook codec modeling with better network designs or larger parameter count.
TS3-Codec~\cite{wu2024ts3codectransformerbasedsimplestreaming} adopts a fully Transformer design that leads to a better single-codebook codec with fewer computation overhead.
LSCodec~\cite{guo2024lscodec} also achieves single-codebook coding with speaker disentanglement (Section \ref{sec:acoustic-disen}).
These single-codebook codecs with remarkably low bitrates offer great benefit to downstream speech generation models on simplicity and efficiency.

\paragraph{Temporal redundancy reduction}

Instead of capturing all the information through VQ layers like the previously mentioned codecs, some researchers have attempted to reduce the redundant bitrate of time-varying VQ codes.
One reasonable method is to encode the global information in speech, e.g. speaker timbre and channel effects, by a global encoder instead of the time-varying codes.
Disen-TF-Codec~\cite{jiang2023disentangled} is the first to explore VQ-GAN codec models with an additional global encoder that aids the codec decoder. 
In Disen-TF-Codec, the global features are designed to be sequential to adapt to speaker changes during transmission.
In TiCodec~\cite{ticodec}, the global tokens are time-invariant and vector-quantized instead.
They are extracted from different segments of an utterance in conjunction with time-varying tokens.
Similar global encoders are also seen in \cite{guo2024socodec,guo2024speaking,singlecodec,wang2025sparktts}.
FreeCodec~\cite{zheng2024freecodecdisentangledneuralspeech} further incorporates a prosody encoder~\cite{ren2022prosospeech} that compresses the low-frequency range of mel spectrograms into a low frame rate VQ sequence to assist in reconstruction.


Another typical example of temporal redundancy reduction is predictive coding, as seen in TF-Codec~\cite{jiang2023latent}.
This approach captures temporal-varying information in the latent space by autoregressive prediction, which significantly reduces redundancy and entropy in the residual part for  quantization.
LMCodec~\cite{LMCodec} employs autoregressive prediction from coarse codes (first RVQ levels) to fine codes (last RVQ levels)~\cite{borsos2023audiolm}, enabling the transmission of fewer codes.

\paragraph{Multi-resolution and variable-bitrate coding}
\label{sec:multi-resolution}

Rather than relying solely on uni-resolution tokens, where all quantizers share the same temporal frequency, it is reasonable to design multi-resolution codecs, because speech contains both fast and slow information streams.
For instance, many vowels exhibit slowly changing characteristics, while events such as explosive consonants and background noises require fine-grained modeling. Therefore, incorporating multiple temporal resolutions in codecs is likely to reduce the necessary bitrate.


SNAC~\cite{Siuzdak_SNAC_Multi-Scale_Neural_2024} is a notable multi-resolution neural speech codec.
It follows the DAC~\cite{kumar2024high} architecture, but in each RVQ layer, residuals are downsampled before codebook look-up and upsampled afterward.
This enables SNAC to have three RVQ streams at a frame rate of 12, 23, 47Hz respectively.
Similarly, CoFi-Codec~\cite{guo2024speaking} achieves multi-resolution tokenization by GVQ quantizers within its U-Net-based architecture.
LLM-Codec~\cite{yang2024uniaudio15} also adopts this idea to achieve very low frame rates with semantic distillation (Section \ref{sec:acoustic-distillation}).

In addition to multiple temporal resolutions, it is also feasible to consider the varying information intensities across different speech frames. 
This observation motivates the design of codecs to allocate different numbers of quantizers for different speech frames.
As an example, VRVQ~\cite{chae2024variable} automatically selects the number of RVQ quantizers per frame by a predictor that is jointly trained with the whole network.

\subsubsection{Challenges}
Despite the emergence of single-codebook and low-bitrate codecs~\cite{singlecodec,ji2024wavtokenizer,xin2024bigcodec,guo2024lscodec}, achieving ideal reconstruction quality with a highly limited VQ space remains a challenging problem. 
Additionally, as acoustic tokens aim to encode all necessary information for signal recovery, they may become redundant and overly complex for downstream modeling.
While scaling up the model size or switching to non-causal networks has been shown to improve performance~\cite{singlecodec,xin2024bigcodec,parker2024scalingtransformerslowbitratehighquality}, these approaches may also compromise streamability or efficiency.
Furthermore, simply introducing global encoders like \cite{jiang2023disentangled,ticodec,guo2024speaking} does not guarantee disentanglement (Section \ref{sec:acoustic-disen}) and may still result in redundancy within the time-varying codes.

\vspace{-0.1in}
\subsection{Acoustic Tokens with Semantic Distillation}
\label{sec:acoustic-distillation}

\begin{figure}
    \centering
    \includegraphics[width=0.99\linewidth]{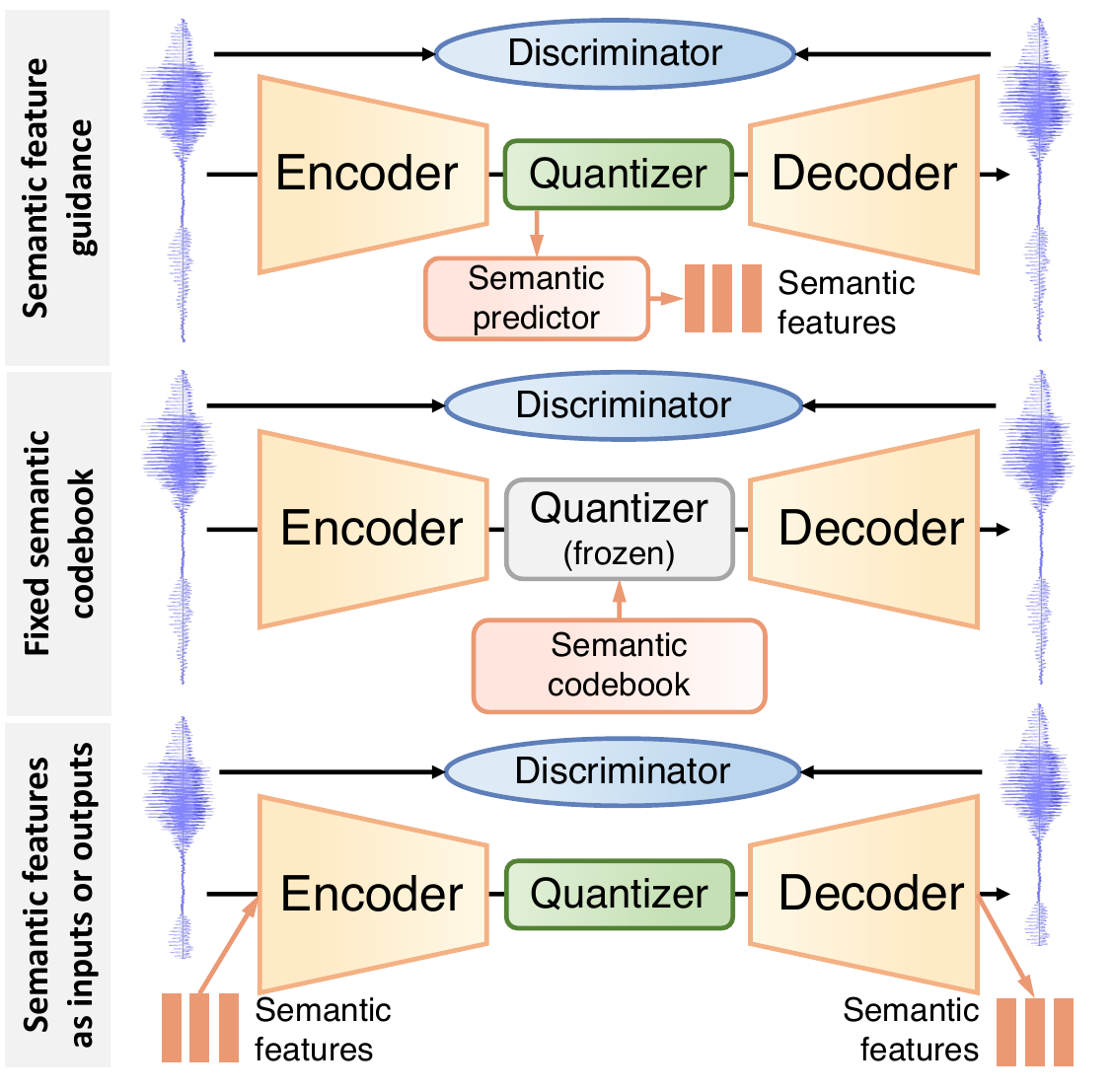}
    \caption{Different semantic distillation methods applied to acoustic tokens (illustrated with the VQ-GAN architecture).}
    \label{fig:acoustic-distill}
    \vspace{-0.2in}
\end{figure}

\subsubsection{Motivation}
Acoustic tokens are a convenient choice for spoken language models, as they can be directly converted back to waveforms without the need for extra vocoders.
However, if reconstruction is the sole objective of these tokens, their representation space may become overly complex and overly focused on acoustic details, in contrast to natural language tokens that primarily carry semantic information.
A natural improvement is to incorporate speech semantic features either from speech self-supervised learning (SSL) models, supervised models, or even text transcriptions.
Since speech SSL models aim to capture high-level phonetic or semantic information without external supervision~\cite{mohamed2022self}, integrating SSL features does not impose additional data requirements for injecting semantic information into the training process. 
Codec models with criteria beyond reconstruction are sometimes referred to as having a ``mixed objective''~\cite{cui2024recent}.
Given that the primary purpose of these models remains acoustic reconstruction in these models, we continue to refer to them as acoustic tokens.
The process of introducing semantic information into acoustic tokens is termed \textbf{semantic distillation}, with approaches summarized in Fig. \ref{fig:acoustic-distill}.

\subsubsection{Approaches}

\paragraph{Semantic feature guidance} 
The earliest effort in semantic distillation is to guide some RVQ layers in codec models towards semantic features, which are typically SSL features. 
Since information in RVQ naturally follows a coarse-to-fine order, guiding early RVQ layers towards semantic-oriented features helps establish and reinforce a semantic-to-acoustic information hierarchy.
For example, SpeechTokenizer~\cite{zhang2024speechtokenizer} uses a HuBERT~\cite{hsu2021hubert} SSL model to guide the first RVQ layer in EnCodec.
This ensures that the first RVQ layer contains more semantic information, thereby pushing acoustic details to the subsequent RVQ layers. 
This distillation is implemented either by regressing the first RVQ output to continuous HuBERT embeddings or by classifying it into discrete HuBERT tokens.
LLM-Codec~\cite{yang2024uniaudio15} alternatively uses Whisper~\cite{whisper} and T5~\cite{raffel2020exploring} as semantic teachers.
Mimi~\cite{kyutai2024moshi} uses a WavLM~\cite{chen2022wavlm} teacher and applies distillation to a specialized VQ module rather than the first RVQ layer.
Paired speech-text data can also be utilized, like in SecoustiCodec~\cite{qiang2025secousticodec} where aligned phoneme sequences serve as the semantic teacher.

Since SSL feature guidance occurs only during the training stage, it does not incur additional inference costs.
It has been reported that language modeling-based TTS trained with such acoustic tokens can exhibit better robustness than those with unguided tokens~\cite{zhang2024speechtokenizer}.

\paragraph{Fixed semantic codebook} A more direct approach to achieve semantic distillation is to integrate semantic knowledge into the codebook of quantizers. 
This forces the quantization space itself to be more semantic-related.
This method is proposed in LLM-Codec~\cite{yang2024uniaudio15} where all three RVQ codebooks are initiated from the token embedding module of LLaMa-2~\cite{touvron2023llama2} and remain frozen during training.
According to \cite{yang2024uniaudio15}, this approach reduces the bitrate and improves the semantic representation ability of LLM-Codec.

\paragraph{Semantic features as inputs or outputs} 
Semantic features can also be compressed together with the acoustic features. 
This requires the encoder and quantizer to construct a shared acoustic and semantic space that balances the two information sources. 
The first attempt in this direction is made in \cite{siahkoohi22_interspeech} where Conformer representations from a pretrained wav2vec 2.0~\cite{baevski2020wav2vec} are combined with CNN encoder outputs for quantization.
SemantiCodec~\cite{liu2024semanticodec} quantizes AudioMAE~\cite{huang2022masked} SSL features
without relying on acoustic inputs. 
The quantized SSL features then serve as a condition for acoustic reconstruction using latent diffusion, which resembles a vocoder that transforms semantic inputs into acoustic outputs.
Providing aligned phoneme sequences instead of SSL features to the quantizer has also shown benefits on reducing bitrates~\cite{du2024funcodec}.

Moreover, semantic features can also serve as outputs, thereby reinforcing the constraint that semantic information be compressed into the discrete latent space.
For instance, \cite{guo2024socodec,ye2024codec} combine hidden HuBERT embeddings with acoustic features before RVQ and jointly optimizes acoustic and semantic reconstruction objectives.
X-Codec 2.0~\cite{ye2025llasa} improves it by using w2v-BERT 2.0~\cite{barrault2023seamless} and FSQ.
XY-Tokenizer~\cite{gong2025xy} further replaces the semantic reconstruction objective by an LLM-based ASR task, aiming at stronger alignment with texts.

\subsubsection{Challenges}
Guiding part of the RVQ layers towards semantic features does not guarantee that acoustic information is only encoded in the remaining layers, as shown by the degraded VC performance in SpeechTokenizer~\cite{zhang2024speechtokenizer}.
It may impose a greater challenge for the VQ layer to encode both acoustic and semantic information if semantic features serve as inputs as well.
Additionally, fixing a semantic codebook could negatively impact acoustic reconstruction ability, as the VQ representation space will become overly restricted.

\vspace{-0.1in}
\subsection{Acoustic Tokens with Disentanglement}
\label{sec:acoustic-disen}
\subsubsection{Motivation}
Another line of mixed-objective acoustic tokens is {disentanglement}.
A prominent research direction is the disentanglement of speaker timbre information, as this is a global trait among all the speech information aspects.
Encoding speaker information into every token timestep is redundant; thus, removing the global speaker timbre can make the information in acoustic tokens more compact and reduce the necessary bitrate. 
Speaker-decoupled speech tokens can alleviate the modeling burden for downstream tasks. For example, a TTS model using these tokens can achieve independent control over prosody and speaker identity.
The disentanglement of speaker timbre also enables an acoustic token to perform voice conversion (VC), as timbre from the target speaker can be easily combined with the speaker-agnostic content tokens from the source speech.

Note that in Section \ref{sec:acoustic}, it is mentioned that some codecs introduce a global encoder to reduce the necessary bitrate of time-variant tokens~\cite{jiang2023disentangled,ticodec,singlecodec,zheng2024freecodecdisentangledneuralspeech}.
They have already demonstrated some ability to decouple global speaker timbre and local contents, albeit in an \textbf{implicit} manner through the natural information bottleneck from VQ.
In this section, we elaborate on \textbf{explicit} methods, which involve specialized training techniques and criteria to achieve disentanglement.
Also, disentanglement is not exclusive with semantic distillation, and many codecs incorporate techniques for both objectives.

\subsubsection{Approaches}
\paragraph{Gradient reversal layer (GRL)} The GRL technique~\cite{drl} is commonly used for disentanglement. Suppose speaker information needs to be disentangled, and a classifier (or speaker verifier, etc.) $s_\mu(\cdot)$ receives some latent feature $\bm h$ from the codec model to perform speaker discriminative tasks. 
GRL operates by negating the gradient sign before $s_\mu(\cdot)$, thereby forcing $\bm h$ to fool the speaker classifier while the classifier itself improves, similar to adversarial training.

SSVC~\cite{SSVC} is one of the pioneering efforts in this direction.
SSVC attempts to decouple content and speaker representations from WavLM features.
The content branch is quantized via RVQ, and the speaker branch is trained using a contrastive loss to produce speaker embeddings.
Disentanglement is enforced by a GRL between the speaker embeddings produced from the speaker branch and the content representations.
Similarly, PromptCodec minimizes an SSIM loss~\cite{wang2004image} between content and speaker representations, with the help of a pretrained speaker verification model.

Such GRL technique is not limited to disentangling speaker timbre alone.
FACodec~\cite{facodec} employs supervised decoupling to factorize speech into speaker timbre, content, prosody, and acoustic detail information.
The timbre extractor in FACodec is optimized via a speaker classification loss.
For the other components -- prosody, content, and acoustic detail -- separate RVQ modules are applied prior to the supervised decoupling process.
For each component, a supervision signal specific to the desired information is applied, and GRL is employed to other non-related information components.
These three quantized features are then combined before applying GRL with the speaker information. 
Finally, the decoder integrates all four information branches to reconstruct the speech signal.

\paragraph{Perturbation}
For speaker disentanglement, a more straightforward approach is to apply speaker timbre perturbations to speech signals and leverage the strong information bottleneck created by the discrete VQ module.
When the encoder is unable to learn sufficient timbre information, and the decoder is provided with prominent timbre, the bottleneck in the middle will naturally prevent timbre from being encoded~\cite{qian2019autovc}.
This idea is adopted in LSCodec~\cite{guo2024lscodec}, which applies a time stretching-based speaker perturbation algorithm to the input waveform.
LSCodec then leverages continuous WavLM features to represent speaker timbre, and feeds them 
to a Conformer-based decoder by position-agnostic cross attention~\cite{du2024unicats,li2024sef}.
Through this approach, LSCodec reports high-quality speech reconstruction and voice conversion using only a single codebook with highly limited bitrates.

\paragraph{Source separation}
Apart from the disentanglement of speaker timbre, source separation has also been explored in the context of acoustic tokens.
SD-Codec~\cite{bie2024learning} and DeCodec~\cite{luo2025decodec} propose to decouple different audio sources in the neural codec by encoding them in separate RVQ modules.
DeCodec also improves separability by enforcing an orthogonality loss.
These approaches allow for more efficient and targeted processing of each audio component, and are also related to the broader scope of universal audio coding.

\subsubsection{Challenges}
The GRL technique for disentanglement inherently carries the risk of a more complex optimization trajectory.
Additionally, some disentanglement methods require supervised data~\cite{facodec}, which imposes a significant constraint.
Due to the intricate nature of speech informatics, current efforts are still suboptimal compared to semantic tokens, particularly in terms of VC performance~\cite{guo2024lscodec}.

\section{Speech Tokenization Methods: Semantic Tokens}

\label{sec:semantic}
Semantic tokens refer to discrete speech representations from discriminative or self-supervised learning (SSL) models.
While we use the term \textit{semantic tokens} to maintain consistency with prior works, some researchers recently argue that SSL features are more accurately described as \textit{phonetic} than \textit{semantic}~\cite{choi24b_interspeech} in nature.
Hence to clarify, in this review, semantic tokens should be more accurately defined as the complementary set of acoustic tokens, such that they are not primarily aimed at reconstruction purposes.
In practice, the vast majority of these tokens are designed for discriminative tasks and are believed to have a strong correlation with phonetic and semantic information~\cite{wells22_interspeech,mohamed2022self,sicherman2023analysing,yeh2024estimating}.

\vspace{-0.1in}
\subsection{Semantic Tokens from SSL Models}
\label{sec:semantic-general}
\subsubsection{Motivation}
When fine-tuned, speech SSL models have shown strong performance on various tasks, often surpassing traditional methods~\cite{superb,mohamed2022self}.
Their potential has been extensively mined in discriminative tasks such as automatic speech recognition (ASR)~\cite{wav2vec,vq-wav2vec,hsu2021hubert,zhang2020pushing}, automatic speaker verification (ASV)~\cite{chen2022wavlm,jung2024espnet,miara24_interspeech}, speech emotion recognition (SER)~\cite{morais2022speech,chen2022wavlm,MADANIAN2023200266,ma-etal-2024-emotion2vec} and speech translation (ST)~\cite{wu20g_interspeech,nguyen20_interspeech,babu22_interspeech}.
Discretized SSL tokens are initially favored for reducing computation costs and improving robustness against irrelevant information for ASR~\cite{chang23b_interspeech}.
Driven by the success of language models, these SSL tokens have been further explored in generative tasks such as TTS~\cite{VQTTS,kharitonov2023speak,vectokspeech} and SLM~\cite{lakhotia2021generative,borsos2023audiolm,hassid2024textually}.
This is because they can be considered high-level abstractions of speech semantics that are largely independent of acoustic details.

\begin{figure}
    \centering
    \includegraphics[width=0.99\linewidth]{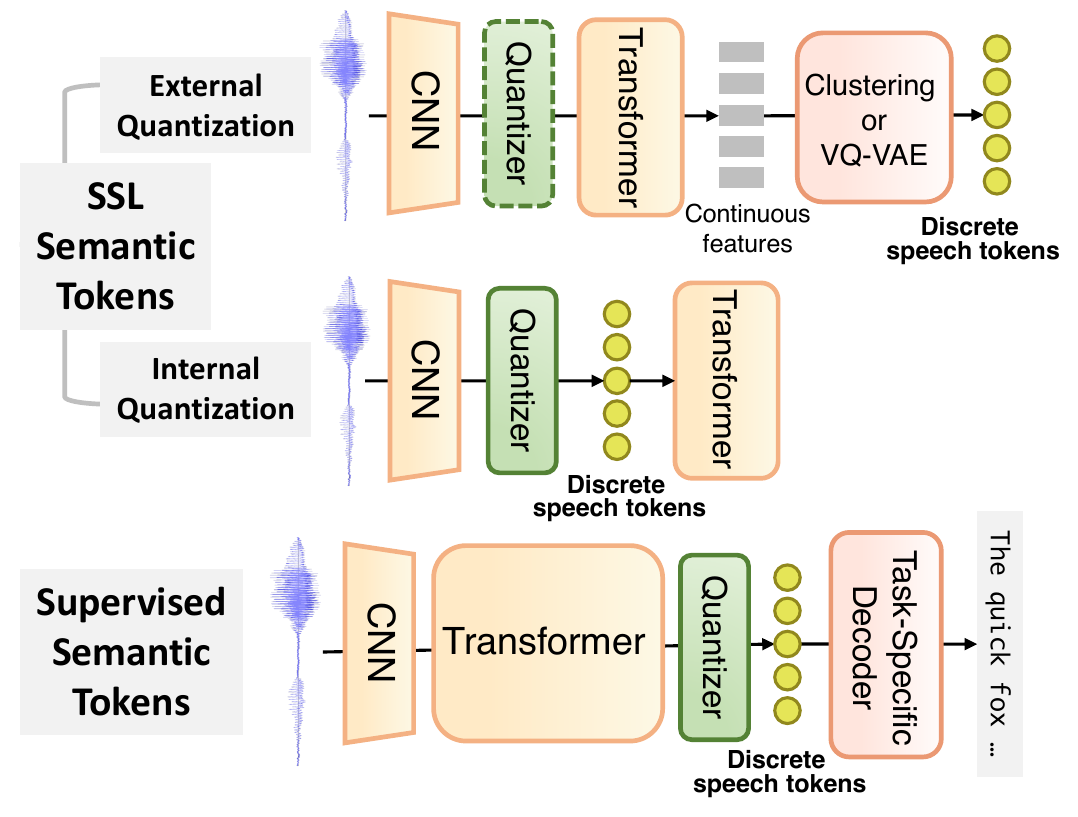}
    \caption{Representatives in different kinds of semantic tokens. 
    Dotted box means the module is optional.}
    \vspace{-0.1in}
    \label{fig:semantic-types}
\end{figure}
\subsubsection{Approaches}

SSL models initiate the learning process by defining a pretext task which enables the model to learn meaningful representations directly from the data itself. 
Typical speech SSL models employ CNNs and Transformer encoders to extract deep contextual embeddings.
When it comes to semantic tokens, there are mainly two ways to extract those discrete tokens from an SSL model (see upper part of Fig.\ref{fig:semantic-types}):
\begin{itemize}[leftmargin=5mm]
    \item External quantization, like clustering or training a VQ-VAE. This refers to extracting continuous embeddings from a certain layer or multiple layers in a pretrained SSL model, and performing quantization manually.
    For example, a common semantic token is the HuBERT+kmeans units, where k-means clustering is performed on a HuBERT Transformer layer with a portion of training data~\cite{lakhotia2021generative,kharitonov-etal-2022-text}.
    It is also feasible to perform clustering on multiple layers~\cite{shi24h_interspeech,mousavi2024should}, or train a VQ-VAE on the SSL hidden embeddings~\cite{huang2023repcodec,wang2024maskgct}.
    \item Internal quantization: when an SSL model contains an inner quantizer that is trained together with other network modules, its outputs can also be regarded as semantic tokens.
    Many SSL models involve quantizers to produce targets for their training objectives~\cite{vq-wav2vec,baevski2020wav2vec,chiu2022self,zhu2025muq}.
    This approach provides an efficient and effective way of extracting semantic tokens.
\end{itemize}
Note that for SSL models with an inner quantizer, it is still practical to perform external quantization on its continuous embeddings, like wav2vec 2.0~\cite{baevski2020wav2vec}.
However, these two methods -- internal and external quantization -- may result in different patterns of information exhibition, which we will investigate in Section \ref{sec:analysis}.

For general-purpose SSL models, there are different designs on the pretext task~\cite{mohamed2022self}.
Table \ref{tab:semantic-metadata}
provides a summary of well-known semantic tokens.

\paragraph{Contrastive models} This type of speech SSL models aims to learn representations by distinguishing a target sample (positives) from distractors (negatives) given an anchor~\cite{mohamed2022self}.
They minimize the latent space similarity of negative pairs and maximize that of the positive pairs.
For semantic tokens, vq-wav2vec~\cite{vq-wav2vec} and wav2vec 2.0~\cite{baevski2020wav2vec} are two representative contrastive SSL models.
They involve a quantizer to produce localized features that is contrastively compared to contextualized continuous features.
Vq-wav2vec~\cite{vq-wav2vec} uses pure CNN blocks while wav2vec 2.0~\cite{baevski2020wav2vec} adopts a Transformer for stronger modeling capacity.
Both use GVQ quantizers to expand the VQ space.
Wav2vec 2.0 has also been extended to massively multilingual versions~\cite{conneau21_interspeech,babu22_interspeech,pratap2024scaling}.

\paragraph{Predictive models}
This type of speech SSL models incorporates an external target for prediction, either from signal processing features or another teacher network.
A popular line of work is HuBERT~\cite{hsu2021hubert}.
It takes raw waveforms as inputs, applies random masks on the hidden representations before Transformer contextual blocks, and then predicts k-means quantized targets from MFCC or another HuBERT teacher.
WavLM~\cite{chen2022wavlm} improves HuBERT by additional speaker and noise augmentations to achieve superior performance in more paralinguistic-related tasks.
There are no inner quantizers in both models, so external quantization like k-means clustering is necessary to obtain semantic tokens.
BEST-RQ~\cite{chiu2022self} changes the prediction target to the output of a random projection quantizer.
The next-token prediction criterion from language models (LMs) have also been adopted into speech SSL~\cite{turetzky2024last,han2024nest}, either with or without a pretrained text LM.
This method emphasizes the autoregressive prediction property of learned tokens that may be better suited for the LM use case.

\paragraph{Perturbation-invariant models}
As SSL tokens feature semantic or phonetic information, a major concern is to improve the resistance against perturbations in the input signal.
This invariance includes noise and speaker aspects that don't affect the contents of speech.
Specifically, speaker-invariant SSL tokens aims to remove speaker information, similar to speaker-disentangled acoustic tokens.
In the training process of perturbation-invariant SSL models, noise~\cite{gat2023augmentation,ccc-wav2vec2.0,messica2024nast,huang2022spiral} or speaker~\cite{qian2022contentvec,chang23_interspeech,chang2024dc,hwang2024removing} augmentations are often explicitly introduced.
Special training losses are then incorporated, like contrastive losses~\cite{huang2022spiral,ccc-wav2vec2.0,qian2022contentvec,hwang2024removing} that distinguish the same spoken content among perturbed distractors, and distribution-based metrics~\cite{chang23_interspeech,messica2024nast,chang2024dc} that minimize the distance of latent features caused by perturbations.

\subsubsection{Challenges}
Firstly, SSL models typically require large amount of data to train, as indicated in 
Table S-II.
For SSL models without a built-in quantizer during pretraining, k-means clustering is a prevalent approach to obtain discrete units.
However, given that most SSL models operate in high-dimensional spaces (e.g., with 768 or 1024 dimensions), the space and time complexity of k-means clustering are substantial. 
The clustering results can sometimes be unreliable due to the curse of dimensionality in Euclidean space.
Moreover, it is often reported, and will also be shown by experiments in Section \ref{sec:analysis}, that discretized SSL units lose much acoustic details after quantization~\cite{polyak21,sicherman2023analysing,mousavi2024dasb}.
Different clustering settings, such as the chosen layer and vocabulary size, can lead to different outcomes within a single model.
Finally, since most SSL models utilize Transformer blocks, their causality and streaming ability are compromised.

While perturbation-invariant SSL models have emerged as promising approaches for semantic tokens, they currently rely on content-preserving augmentations that are typically hand-crafted.
Most methods in this type have only been evaluated on small-scale data and models.
It also remains unclear how these methods will generally benefit generative tasks such as speech generation and spoken language modeling.

\IEEEpubidadjcol

\subsection{Semantic Tokens from Supervised  Models}
\label{sec:semantic-supervised}
As representing semantic or phonetic information is the major purpose of semantic tokens, a more direct way to achieve this is through supervised learning.
Supervised semantic tokenizers are typically trained on the ASR task.
A famous example shown at the bottom of Fig.\ref{fig:semantic-types} is the $\mathcal S^3$ Tokenizer from CosyVoice~\cite{du2024cosyvoice}.
It places a single-codebook VQ layer between two Transformer encoder modules and optimizes the network using a cross-entropy ASR loss, similar to Whisper~\cite{whisper}.
The same method is adopted in \cite{zeng2024scaling,zeng2024glm} where the frame rate is further reduced to 12.5Hz.
CosyVoice 2~\cite{cosyvoice2} improves $\mathcal S^3$ Tokenizer by replacing plain VQ with FSQ for better codebook utilization.
Note that in this kind of supervised semantic tokens, it is the output of the VQ layer that serves as tokens.
This allows for more preservation of paralinguistic information than directly transcribing speech into text.
CosyVoice 3~\cite{du2025cosyvoice3} extends supervised tokens to more tasks involving language, emotion, speaker and audio analysis.
These supervised tokenizers are trained on massive paired speech-text data, and have demonstrated rich speech content understanding capabilities~\cite{du2024cosyvoice,fang2024llamaomni}.

However, training these models is highly costly due to the heavy data demands.
Training with only the ASR task may still result in the loss of some prosody information, and training with multiple tasks poses a higher demand for data and task balancing.
Although \cite{cosyvoice2} has demonstrated that its supervised tokenizer trained on Chinese and English can also work in Japanese and Korean, it remains unclear how well these supervised tokenizers generalize to more unseen languages.

\subsection{Speech Token Vocoders}
\label{sec:vocoder}
Acoustic tokens are naturally coupled with a decoder that outputs waveforms or spectrograms given tokens, but semantic tokens do not inherently require a vocoder, particularly for speech understanding tasks.
However, when semantic tokens are used for speech generation, a speech token vocoder (also known as speech resynthesis model) becomes necessary.
Unlike traditional spectrogram-based vocoders~\cite{kong2020hifigan}, speech token vocoders  usually need to compensate the loss of acoustic details in semantic tokens by introducing additional inputs.

Polyak et al.~\cite{polyak21} first explores speech resynthesis from discrete speech units by a HifiGAN~\cite{kong2020hifigan} augmented with discretized pitch units and speaker embedding inputs.
\IEEEpubidadjcol
The vec2wav vocoder in VQTTS~\cite{VQTTS} improves this vocoder by a Conformer~\cite{conformer} frontend module before HifiGAN generator.
Later, CTX-vec2wav~\cite{du2024unicats} proposes a position-agnostic cross-attention mechanism that effectively integrates timbre information from surrounding acoustic contexts without the need for pretrained speaker embeddings.
This makes it more timbre-controllable and suitable for zero-shot TTS and VC~\cite{li2024sef}.
Upon it, vec2wav 2.0~\cite{guo2024vec2wav} introduces SSL timbre features and adaptive activations to improve timbre controllability, and reports competitive VC performance.

It is also feasible to apply diffusion or flow matching algorithms in token vocoders~\cite{tortoise,seedtts,du2024cosyvoice}.
There, the discrete tokens are treated as a condition for diffusion or flow matching to generate mel-spectrograms, and further converted to waveform by a pretrained mel vocoder.
Compared to training a token-to-wav vocoder in an end-to-end fashion, training a token-to-mel model is more convenient and does not need adversarial training. 
To better control timbre, a mask strategy is introduced into the training process where the model only computes loss on the un-masked part of spectrograms~\cite{du2024cosyvoice}.
During inference, spectrogram from speaker prompt conditions the generative process, which can be regarded as a form of ``in-context learning''.
However, this requires tokens to be extracted from reference prompts before synthesis.
Also, inference efficiency may be compromised for better generation quality with multiple inference steps, and this method is only validated on massive amount of data currently.

\section{Length Reduction and Variable-Rate Tokenization}
\label{sec:length-and-vfr}

In most cases, the frame rate of discrete speech tokens ranges from 25 to 100Hz.
This leads to a huge discrepancy in lengths between speech representations and the underlying text modality.
This discrepancy has been a critical issue in building decoder-only TTS and other LM-based speech generation tasks~\cite{wang2024whyspeech}, since longer sequences result in harder training and more unstable inference.
Therefore, researchers have proposed different ways to mitigate this issue, either by post-processing on tokens, or learning-based variable-rate tokenization.

\subsection{Length Reduction by Deduplication and Acoustic BPE}
\label{sec:dedup-bpe}

Post-training length reduction methods for speech tokens are inspired by language processing techniques. 
A common approach to reduce token sequence lengths is deduplication~\cite{chang23b_interspeech,chang2024exploring}, i.e. removing the repeated consecutive tokens in a sequence.
Since the encoded continuous features are often close in consecutive frames where the speech dynamics do not change rapidly, they are likely to be quantized to the same unit.
Therefore, removing these redundant tokens can yield a more phonetic representation.
When the deduplicated tokens are used for generative modeling, a unit-to-speech model (similar to TTS) should be employed to upsample the tokens and convert them back to acoustic signals~\cite{lakhotia2021generative}.

Another popular approach is acoustic byte-pair encoding (BPE)\footnote{The term ``acoustic'' here is used to distinguish it from traditional BPE applied to text tokens, rather than referring to ``acoustic tokens''.} or so-called subword modeling~\cite{hayashi2020discretalk,ren22_interspeech,chang23b_interspeech,shen2024acoustic,dekel24_interspeech}.
Similar to text BPE~\cite{Gage1994ANA}, acoustic BPE iteratively merges the two most frequent consecutive tokens and adds the merged token to the vocabulary.
After training on a corpus, a deterministic BPE mapping is established between original token combinations and the new vocabulary. 
This mapping enables a lossless compression algorithm, allowing tokens to be perfectly reconstructed after BPE decoding.
This operation can identify certain morphological patterns in token sequences, and offers a powerful way to remove redundant tokens.
In practice, acoustic BPEs on HuBERT semantic tokens has demonstrated significant speed and performance gains in ASR~\cite{chang23b_interspeech,chang2024exploring}, spoken language modeling~\cite{shen2024acoustic,dekel24_interspeech} and TTS~\cite{li24qa_interspeech,vectokspeech}.

Although deduplication is a simple and training-free method, acoustic BPE offers several unique advantages over it. First, acoustic BPE can identify redundant patterns that are not simply repetitions, whereas deduplication only removes exact duplicates. Second, deduplication discards the duration information of every token in the resulting sequence. This could be problematic for downstream tasks, as important rhythmic information may reside in the repetitions of tokens. In contrast, acoustic BPE preserves duration information by encoding repetitions of varying lengths into distinct new tokens. Third, acoustic BPE is more flexible than deduplication in terms of target vocabulary size, which can be adjusted based on the desired length reduction ratio and downstream performance.

\begin{figure}
    \centering
    \includegraphics[width=0.99\linewidth]{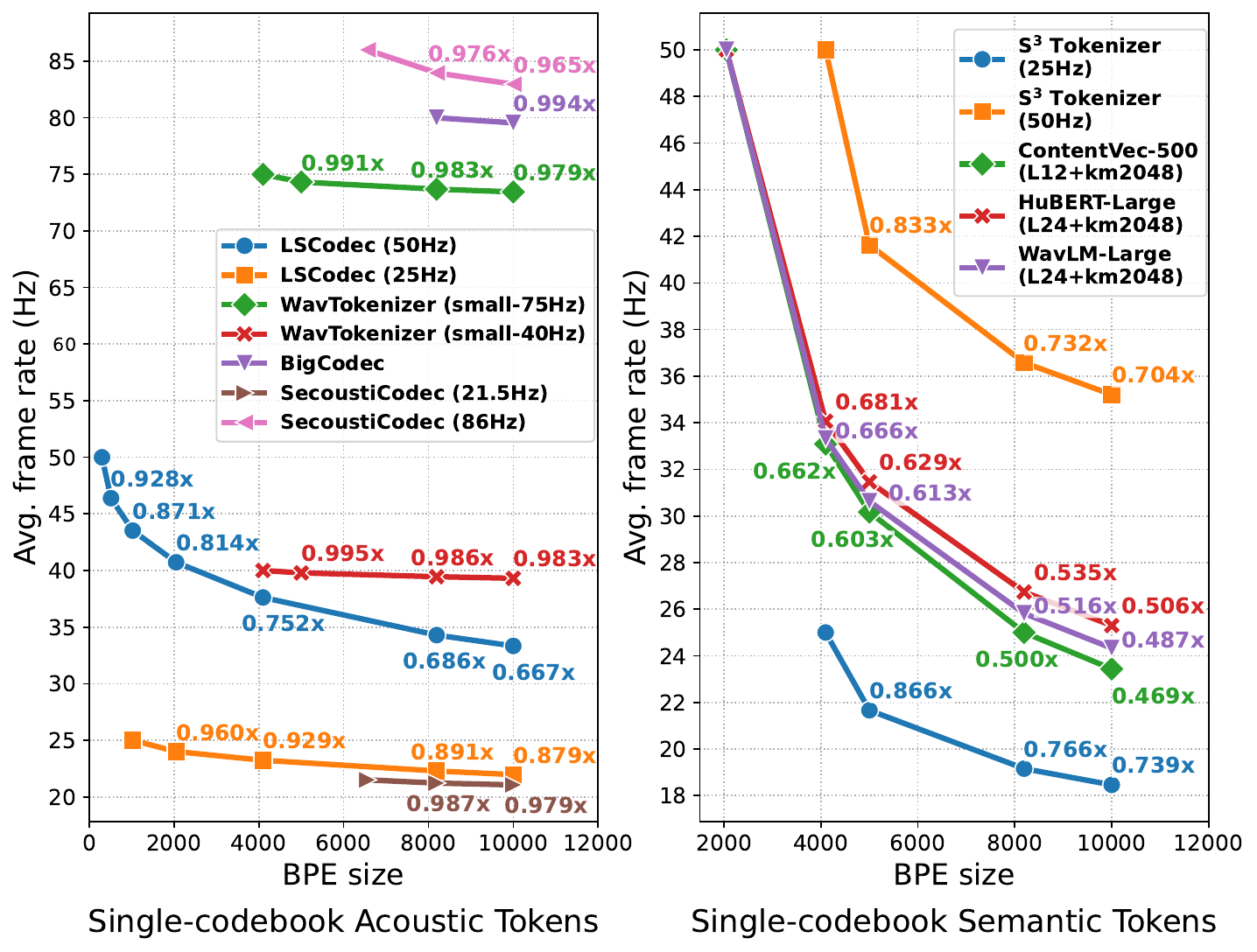}
    \caption{BPE effect comparison of multiple tokens. The starting point of each line represents the original vocabulary size.}
    \vspace{-0.15in}
    \label{fig:bpe-effect}
\end{figure}
We visualize the length reduction effect of BPE on different speech tokens in Fig.\ref{fig:bpe-effect}, including acoustic and semantic tokens all with a single codebook. 
From Fig.\ref{fig:bpe-effect}, it is evident that different types of tokens exhibit very distinct patterns.
Semantic tokens generally show significant length reduction when applying BPE.
For single-codebook acoustic tokens, speaker-decoupled LSCodec tokens show more reduction than general-purpose WavTokenizer and BigCodec.
This suggests that the effect of BPE is negatively correlated with the information density in the speech tokens: the less information they contain, the more length reduction is achieved by BPE.

\subsection{Variable Frame Rate Tokens and Unit Discovery}
\label{sec:variable-rate}
Information in speech is not uniformly distributed along the time axis~\cite{dieleman2021variable}.
In segments such as silence or long vowels, information density is low, whereas in segments with explosive consonants, speech events occur much more frequently.
This inherent non-uniformity suggests that it might be more natural to allocate more tokenized bits to regions with dense information and higher variance, and fewer bits to regions with less uncertainty.
This kind of discrete speech tokens is referred to as \textit{variable frame rate (VFR) tokens} in this review.
Note that while multi-resolution and variable-bitrate tokens have been introduced previously, the concept of VFR is still distinct.
In multi-resolution tokens~\cite{Siuzdak_SNAC_Multi-Scale_Neural_2024,guo2024speaking}, each quantizer operates at a fixed frame rate.
In variable-bitrate tokens~\cite{chae2024variable}, the frame rate remains fixed, while the variability lies in the  number of quantizers per frame.
Instead, VFR tokens should directly allocate different granularities on the temporal axis.

VFR tokens are closely related to acoustic unit discovery. As speech lacks a natural boundary of phonetic units~\cite{mohamed2022self}, there are much research efforts to find and locate the underlying acoustic units behind speech utterances in an unsupervised manner~\cite{eloff19_interspeech,dunbar20_interspeech,niekerk20b_interspeech,nguyen2020zero}.
This is particularly of interest for low-resource languages.
The discovered units can guide the boundary segmentation of VFR tokens.
To this end, VFR tokens are interesting not only because they might reduce the necessary bitrate, but also because they can introduce a strong inductive bias that linguistic knowledge is encoded~\cite{dieleman2021variable}.

A recent direction of VFR tokens is to discover acoustic units from an SSL model.
Sylber~\cite{cho2024sylber} and SyllableLM~\cite{baade2024syllablelm} take similar approaches that first heuristically locate acoustic boundaries from existing HuBERT models, and then train another HuBERT student with segment-level pooled targets between boundaries.
The final HuBERT embeddings undergo the same segment-level pooling and kmeans clustering procedure to produce tokens at a very low frame rate ($\approx5$Hz).
Such tokens are reported to align well with syllables~\cite{baade2024syllablelm,cho2024sylber}.

Boundary prediction can be involved to achieve frame rate variability in the training process, where a specific model predicts frame-level boundaries and is trained together with other network modules.
The training techniques of such models include reinforcement learning~\cite{cuervo2022variable}, soft pooling~\cite{hwang2024removing}, and slowness constraint~\cite{dieleman2021variable}.
For VFR acoustic tokens, heuristic downsampling methods have been explored recently.
TFC~\cite{zhang25k_interspeech_tfc} incorporates a frame-rate selection strategy based on entropy values, assigning different downsampling rates to segments with varying entropy levels under a shared quantizer. CodecSlime~\cite{wang2025codecslime} adopts a dynamic programming-based downsampling strategy driven by pairwise latent embedding distances, and further introduces a specialized training method to adapt fixed frame-rate codec models to dynamic frame rates. Both approaches enable flexible frame-rate allocation within a single model, and are orthogonal to the codec architecture. 
However, the study of VFR acoustic tokens remains generally insufficient.

Similar to supervised semantic tokens, explicit linguistic boundaries can also be incorporated into VFR design.
Karapiperis et al.~\cite{sotirios2024investigating} and TASTE~\cite{tseng2025taste} are an initial explorations in this direction, where the tokens are aligned with the frame rate of phonemes or text BPE units, resulting in variable temporal spans. 
Karapiperis et al.~\cite{sotirios2024investigating} is an acoustic token that directly performs phoneme-level downsampling before quantization.
In contrast, TASTE aggregates information from a whole spoken utterance onto each text BPE unit by cross-attention.
The BPE-level aggregated speech embeddings are then quantized and optimized using a TTS-style objective,
These works redefine the notion of tokens: rather than encoding spoken information by themselves, they serve as an additional acoustic layer conditioned on explicit text content.


\section{Analysis of Discrete Speech Tokens}
\label{sec:analysis}
\subsection{Metrics and Existing Benchmarks}

Discrete speech tokens can be evaluated from various aspects besides bitrate and codebook utilization:
\begin{itemize}
    \item \textbf{Signal-level reconstruction metrics}: 
    For reconstruction evaluations, signal-level metrics like
    PESQ~\cite{rix2001perceptual}, STOI~\cite{taal2011algorithm}, mel distance, GPE~\cite{bagshaw1993}, etc. are often used.
    \item \textbf{Perceptual reconstruction metrics}: Apart from  signal-level metrics, there can also be perceptual evaluations of reconstruction performance. This includes intelligibility (often measured by word, character, or phone error rates), speaker similarity, subjective listening tests, etc.
    \item \textbf{Performance on downstream tasks}: Probing tasks can be used to measure the preservation or prominence of certain information in tokens, like ASR, ASV, emotion recognition, and spoken language modeling~\cite{nguyen2020zero}.
    Note that this is different from perceptual reconstruction metrics since it operates directly on tokens.
    Performance in generative tasks like TTS and VC can also be evaluated.
    \item \textbf{Semantic/phonetic relevance}: If the tokens are expected to align with texts (e.g. for semantic tokens and semantic-distilled acoustic tokens), metrics like phone discrminability~\cite{nguyen2020zero}, phone purity~\cite{hsu2021hubert}, and phone-normalized mutual information~\cite{hsu2021hubert} can be computed.
    \item \textbf{Invariance and robustness}: If the tokens are expected to be invariant to perturbations, unit edit distance~\cite{gat2023augmentation} can be considered as a measurement.
\end{itemize}
There are several existing benchmarks on discrete speech tokens. 
Codec-SUPERB~\cite{wu2024codec}
evaluates both signal-level reconstruction metrics and downstream performances of acoustic tokens.
ESPnet-Codec~\cite{shi2024espnet} integrates multiple codecs into a unified training and evaluation framework VERSA~\cite{shi2024versa} and extends evaluation to some generative tasks such as TTS and SVS.
DASB~\cite{mousavi2024dasb} 
performs more downstream probing tasks, and includes generative tasks as well as semantic tokens.
STAB~\cite{vashishth2024stab} takes a different perspective that measures the invariance, robustness, compressibility, and vocabulary utilization of speech tokens. 
This emphasizes the application in spoken language models instead of reconstruction.

\vspace{-0.1in}
\subsection{Existing Analyses}

There are several theoretical or experimental analyses of the advantages of discrete speech tokens.
Nguyen et al.~\cite{nguyen2022discrete} demonstrates by an encoder-only language model that semantic SSL tokens are favorable for spoken language modeling, due to their removal of linguistically irrelevant information.
Sicherman et al.~\cite{sicherman2023analysing} supports this claim by showing that semantic units have a strong correlation with phonemes, but a weaker correlation with speakers.
Abdullah et al.~\cite{abdullah23_interspeech} refines the correlation between semantic SSL tokens and linguistic content to the ``sub-phonemic'' level instead of high-level phonemes due to contextual variability.
Chang et al.~\cite{chang23b_interspeech} explores the use of WavLM tokens for end-to-end ASR together with deduplication and BPE. Although these tokens underperform continuous SSL features, they still show competitive performance. 
Similar findings are reported on contextual ASR~\cite{cui2024exploring_context}, multilingual ASR~\cite{cui2024exploring}, end-to-end speech translation, 
understanding~\cite{chang2024exploring}, and more LLM-based semantic-related tasks with discrete units as inputs~\cite{wang2024comparative}.
Expresso~\cite{expresso}  evaluates the resynthesis quality of HuBERT and EnCodec tokens on an expressive dataset, finding that HuBERT struggles to preserve source speech expressivity while EnCodec performs better.

The downside of speech tokens is also studied.
Yeh et al.~\cite{yeh2024estimating} suggests that VQ on HuBERT embeddings cannot achieve perfect disentanglement of speaker and phonetic content. 
EMO-Codec~\cite{ren2024emo} shows that codec reconstruction still sometimes degrades the emotion information.
O'Reilly et al.~\cite{o2024code} shows that neural audio codecs often lack stability after repeated encoding and decoding, i.e. not idempotent.
ERSB~\cite{wang2025towards-asru} reveals speech codecs still struggle under complex acoustic environments, in terms of both reconstruction and downstream task consistency.

Therefore, the reconstruction quality of acoustic tokens and the performance on discriminative downstream tasks of both acoustic and semantic tokens have been benchmarked.
However, the reconstruction performance of semantic tokens still requires a more thorough comparison.
Hence, we adopt a reconstruction approach to compare different types of tokens.
Specifically, we use a timbre-controllable speech token vocoder to resynthesize semantic tokens into waveforms and measure the preservation of content, prosody, speaker identity, and acoustic details, respectively. 
We also follow DASB~\cite{mousavi2024dasb} to conduct semantic modeling probing tasks on various tokens.
The detailed setups of these experiments can be found in 
the appendix.

\IEEEpubidadjcol

\vspace{-0.15in}
\subsection{Reconstruction Analysis}
\label{sec:reconstruction}

\begin{table*}[]
\centering
\caption{Reconstruction, voice conversion, and downstream semantic modeling comparisons of tokens in different categories.
For reconstruction and voice conversion, we train a specific CTX-vec2wav$^\alpha$ vocoder~\cite{guo2024lscodec} on LibriTTS~\cite{libritts} for all semantic tokens.
Settings in parentheses denote model versions.
Semantic modeling tasks, ASR and IC (intent classification), follow the DASB~\cite{mousavi2024dasb} setup which uses token indices to train a simple LSTM network.
``L'' means a certain layer in the SSL Transformer block, ``km'' means manual k-means clustering, and ``\textit{NC}'' is short for ``not converged''.
Please refer to the appendix for evaluation details.
}
\label{tab:recon-vc}
\resizebox{2\columnwidth}{!}{
\begin{tabular}{@{}llccccccccccc@{}}
\toprule
\multicolumn{1}{c}{\multirow{2}{*}{\textbf{Token Type}}} & \multicolumn{1}{c}{\multirow{2}{*}{\makecell{\textbf{Token}\\Model (version)}}} & \multicolumn{1}{c}{\multirow{2}{*}{\makecell{\textbf{Bitrate$\downarrow$}\\(kbps)}}} & \multicolumn{4}{c}{\textbf{Reconstruction}} & \multicolumn{3}{c}{\textbf{Voice Conversion}} 
& \multicolumn{2}{c}{\textbf{Semantic Modeling}}
\\ \cmidrule(l){4-7}\cmidrule(l){8-10} \cmidrule(l){11-12} 
\multicolumn{1}{c}{} & & \multicolumn{1}{c}{} & \multicolumn{1}{c}{\textbf{WER$\downarrow$}} & \multicolumn{1}{c}{\textbf{GPE$\downarrow$}} & \multicolumn{1}{c}{\textbf{PESQ$\uparrow$}} & \multicolumn{1}{c}{\textbf{STOI$\uparrow$}} & \multicolumn{1}{c}{\textbf{WER$\downarrow$}} & \multicolumn{1}{c}{\textbf{SECS$\uparrow$}} & \multicolumn{1}{c}{\textbf{P.Corr$\uparrow$}} 
& \multicolumn{1}{c}{\textbf{ASR WER}$\downarrow$} 
& \multicolumn{1}{c}{\textbf{IC ACC}$\uparrow$}
\\ 
\midrule
\multirow{2}{*}{\textbf{Continuous Baselines}}  & Ground Truth Recording & - & 1.16 & 0.00 & 4.50 & 1.000 &
- & - & - & - & -\\
 & Mel + BigVGAN (100 band) & - & 1.18 & 0.88 & 4.30 & 0.995 & 
 - & - & - & 17.4 & 50.1 \\
\midrule
\multirow{6}{*}{\makecell[l]{\textbf{Acoustic} \\ (General-Purpose)}} & EnCodec ($Q=8$) & 6.00 & 1.53 & 1.33 & 2.83 & 0.946 & 
- & - & - & 19.4 & 34.8\\
 & DAC (24kHz, $Q=8$) & 6.00 & 1.34 & 0.93  & 3.52 & 0.958 & 
 - & - &-  & 26.1 & 18.3 \\
 & TiCodec ($Q=2$) & 1.50 & 3.31 & 1.51 & 2.00 & 0.898 & 
 2.62 & 0.642 & 0.886  & 26.4 & 41.7
 \\
 & SNAC (24kHz) & 0.98 & 2.25 & 1.48 & 2.23 & 0.914 & 
 - & - & - & 35.4 & 16.7\\
 & WavTokenizer (Small, $F=75$Hz) & 0.90 & 2.45 & 1.63 & 2.47 & 0.925 & 
 - & - & - & 37.2 & 15.5\\
 & Stable-Codec (2 residual FSQ) & 0.70 & 4.94 & 1.73 & 2.16 & 0.917 & 
 - & - & - & \textit{NC} & 14.0 \\
 \cmidrule(r){1-2}
\multirow{5}{*}{\makecell[l]{\textbf{Acoustic}\\(Semantic Distillation)}} & SpeechTokenizer & 4.00 & 1.47 & 1.20 & 2.60 & 0.930 & 
- & - & - &  19.3 & 57.3 \\
& X-Codec (HuBERT LibriSpeech) & 4.00 & 1.27 & 1.49 & 2.82 & 0.905 & 
- & - & - & 9.8 & 69.6\\
& Mimi & 1.10 & 2.44 & 1.68 & 2.27 & 0.917 & 
- & - & - & 26.8 & 50.9\\
& LLM-Codec & 0.85 & 6.25 & 1.86 & 1.82 & 0.879 & 
- & - & - & \textit{NC} & 16.2 \\
& SemantiCodec ($F$=25Hz, $V$=$2^{13}$+$2^{15}$) & 0.70 & 3.44 & 2.28 & 1.75 & 0.866 &  
- & - & - & \textit{NC} & 26.8\\
\cmidrule(r){1-2}
\multirow{2}{*}{\makecell[l]{\textbf{Acoustic}\\(Disentanglement)}} & {FACodec (with \textit{detail} codes)} & 4.80 & 1.37 & 1.02 & 2.91  & 0.954 & 
1.57 & 0.773 & 0.583 & 14.6 & 51.1
\\ 
 & LSCodec ($F=50$Hz) & 0.45 & 3.33 & 2.42 & 1.77 & 0.688 & 
 4.04 & 0.852 & 0.697 & 25.3 & 49.8
 \\ 
\cmidrule(r){1-2}
\multirow{6}{*}{\textbf{Semantic} (SSL)} & vq-wav2vec (k-means) & 1.80 & 2.81  & 2.73 & 1.49 & 0.795 & 
3.27 & 0.857 & 0.718 & 16.9 & 58.7
\\
 & wav2vec 2.0 (Large, inner quantizer) & 0.90 & 3.24 & 2.92 & 1.52 & 0.680 & 
 4.40 & 0.814 & 0.759  & 22.0 & 51.9
 \\
 & wav2vec 2.0 (Large, L14+km2048) & 0.55 & 2.51 & 9.57 & 1.20 & 0.630 & 
 2.81 & 0.880 & 0.492 & 5.8 & 69.5
 \\
 & HuBERT (Large, L24+km2048) & 0.55 &  1.86 & 15.65 & 1.17 & 0.625 & 
 1.97 & 0.876 & 0.375 & 6.1 & 67.2
 \\
 & WavLM (Large, L24+km2048) & 0.55 & 1.67 & 17.94 & 1.16 & 0.621 & 
 1.92 & 0.872 & 0.374 & 6.1 & 74.2
 \\
& ContentVec (L12+km2048) & 0.55 &2.09  & 18.88 &  1.15 &  0.613 & 
2.21 & 0.869 & 0.348 & 5.5 & 72.0
\\ 
\cmidrule(r){1-2}
\textbf{Semantic} (Supervised) & $\mathcal S^3$ Tokenizer ($F=50$Hz) & 0.60 & 2.12 & 4.25 & 1.37 & 0.673 & 
2.52 & 0.868 & 0.687 & 17.5 & 67.2
\\
\bottomrule
\end{tabular}
}
\end{table*}

To enable a fair comparison between acoustic and semantic tokens from a reconstruction perspective, we train a CTX-vec2wav$^\alpha$ vocoder~\cite{guo2024lscodec} for different semantic tokens on LibriTTS~\cite{libritts}.
This vocoder supplements the insufficient speaker timbre information in semantic tokens using continuous WavLM features extracted from the reference prompts.
This approach enables semantic tokens to perform voice conversion (VC) by switching reference prompts conveniently.
The training details follow \cite{du2024unicats}.
We compute several metrics for reconstruction ability:
\begin{itemize}
    \item \textbf{WER} (word error rate, in percentage) measures the content intelligibility of reconstructed speech. It is computed between ground truth texts and ASR-decoded transcriptions. We use NeMo-ASR\footnote{\scriptsize\url{https://huggingface.co/nvidia/stt_en_fastconformer_transducer_large}} here.
    \item \textbf{GPE} (gross pitch error, in percentage) measures the relative error percentage of pitch contours of the reconstructed speech compared to ground truth.
    \item \textbf{PESQ} (perceptual evaluation of speech quality) and \textbf{STOI} (short-time objective intelligibility) measure the speech quality from a signal perspective.
\end{itemize}
We use LibriTTS testset-B~ \cite{du2024unicats} as the test set for evaluations.
It contains 500 utterances from unseen speakers that sum up to about 1 hour.
We use the original utterance to provide timbre information when necessary, i.e. for TiCodec, FACodec, LSCodec and all semantic tokens.
All evaluation metrics are computed on 16kHz waveform for a fair comparison, and reconstructed waveforms with higher sampling rates are downsampled before evaluation.

We take representative works in each token category. 
When there are multiple feasible configurations for a model, we choose one typical configuration that balances bitrate and performance.
Note that different variants (especially on frame rate and number of quantizers) within the same model can lead to significant differences in reported metrics.
For SSL models like wav2vec 2.0, HuBERT and WavLM, we take the official ``Large'' model variant.
For wav2vec 2.0, we experiment with both its inner quantizer before the Transformer blocks and k-means clustering results on a specific Transformer layer.

The results are shown in Table \ref{tab:recon-vc}.
It is evident that acoustic tokens designed only for reconstruction can achieve decent speech quality, but still far from the state-of-the-art spectrogram-based vocoders because of higher compression rates.
Retaining good speech intelligibility (i.e. low WER) becomes particularly challenging when the frame rate is low. 
Acoustic tokens with semantic distillation can also achieve strong reconstruction quality.
Explicitly disentangled acoustic tokens may sacrifice some reconstruction performance metrics when the bitrate is extremely low.
Semantic tokens generally struggle to achieve the same level of acoustic reconstruction as acoustic tokens, as evidenced by lower GPE, PESQ, and STOI scores.
Notably, most semantic tokens included exhibit significant information loss in prosody as reflected by their GPE scores.
However, their WER scores remain comparable to acoustic tokens, despite having much lower bitrates.
This highlights the property that semantic tokens primarily retain content-related information rather than acoustic details.

\subsection{Voice Conversion Analysis}
Despite the loss of acoustic information, a prominent advantage of semantic tokens over most acoustic tokens is their inherent timbre controllability.
Some acoustic tokens also have this ability, such as those with a global encoder like TiCodec and disentangled acoustic tokens, also possess this ability
To compare this ability across different tokens, we conduct voice conversion (VC) experiments using these tokens as the content from the source speech. 
We use the same source utterances in Section \ref{sec:reconstruction}, but assign a different target speaker for each source utterance as the prompt.

Then, we perform VC experiments on the 500 VC pairs.
In addition to WER, we also measure \textbf{SECS} (speaker embedding cosine similarity)~\cite{casanova2022yourtts} as the metric for speaker similarity, and \textbf{P.Corr}~\cite{guo2024vec2wav} as an objective metric for prosody preservation. 
SECS requires a speaker verification model\footnote{\scriptsize\url{https://github.com/resemble-ai/Resemblyzer}} to output speaker embeddings.
P.Corr calculates the Pearson correlation coefficient between the pitch contours of the converted and source utterances. 
Note that P.Corr will be meaninglessly high if the VC similarity is low, i.e., when the source timbre is barely changed.
As the source utterances are the same as Section \ref{sec:reconstruction}, the WER numbers are directly comparable to those in the reconstruction experiments.

The results presented in Table \ref{tab:recon-vc} indicate that semantic tokens often achieve much higher VC similarity compared to acoustic tokens. 
However, due to the substantial loss of prosody information, semantic tokens tend to have lower P.Corr scores than acoustic tokens.
Among the acoustic tokens capable of performing VC, explicit disentanglement methods, such as FACodec and LSCodec, outperform the implicit criterion employed in TiCodec.
It is also noteworthy that different tokenization settings in wav2vec 2.0 lead to drastically different outcomes. 
Tokens generated from its inner quantizer preserve prosody well but also retain much speaker information, whereas clusters derived from its Transformer hidden embeddings exhibit the opposite characteristics.

Supervised semantic tokens from $\mathcal S^3$ Tokenizer also exhibit good intelligibility and VC ability.
Unlike HuBERT-style SSL models, this supervised tokenizer demonstrates better preservation of prosody both in reconstruction and VC settings.
Given that prosody and intonation are a crucial factors for ASR, it is reasonable to assume that the tokenizer's VQ module encodes some prosody information.
In contrast, while HuBERT-style SSL models do contain rich prosody information in their continuous features (e.g., as evidenced by good emotion recognition results~\cite{superb}), phonetic information is likely the primary component.
Therefore, offline clustering is prone to discard these prosody characteristics.

\subsection{Downstream Semantic Modeling Analysis}
Besides reconstruction and voice conversion which mainly cover acoustic aspects, we also explore the semantic modeling abilities in the tokens.
To this end, we follow the DASB~\cite{mousavi2024dasb} setup and considers two representative downstream semantic modeling tasks: ASR and intent classification (IC).
In both tasks, we train a small LSTM-based probing network to process the discrete speech tokens and produce outputs.
IC is a classification task to determine the spoken intents from speech directly, where we use the SLURP~\cite{bastianelli-etal-2020-slurp} dataset with 18 target classes and accuracy (ACC) as the metric.
For ASR, we use LibriSpeech~\cite{librispeech} 960h to train the probing network and report WER on test-clean split.

The results can be found at the last two columns of Table \ref{tab:recon-vc}.
General-purpose acoustic tokens show relatively worse semantic modeling abilities than others under a small probing model, although some acoustic tokens have excellent reconstruction quality.
Semantic distillation can enhance such downstream performance, especially X-Codec which uses SSL features both as input and outputs.
Meanwhile, semantic tokens generally achieve significantly better results in such tasks, with predictive SSL tokens being the best.
For semantic tokens, downstream ASR performance appears to be negatively correlated with prosody preservation: tokens that better preserve prosody often yield worse ASR results. 
This trend is observed in vq-wav2vec, wav2vec 2.0, and even the supervised $\mathcal{S}^3$ Tokenizer, all of which employ an internal quantizer.
In summary, these experiments reveal a trade-off between how strongly semantic information is emphasized in the tokens, and how comprehensively it is preserved.


\IEEEpubidadjcol

\section{Discrete Speech Token-based Application Paradigms}
\label{sec:application}

\subsection{Spoken Language Understanding}
\subsubsection{Motivation}
Spoken language understanding (SLU) tasks, including automatic speech recognition (ASR), speech translation, intent classification and others, aim to extract meaningful domain-specific information from speech.
Most SLU tasks follow a speech-in text-out pipeline, except S2ST which also involves speech generation.
The adoption of discrete tokens in SLU offers some benefits. 
Discrete tokens may naturally exhibit some invariance against noise and speaker information, particularly semantic tokens, which can make downstream models to focus more effectively on content-related information in some tasks.
On a broader scale, discrete tokens provide a promising approach to unifying speech understanding and generation in spoken language models.

As an alternative input to an SLU model instead of continuous features, discrete speech tokens are typically deduplicated or BPE-encoded before subsequent modules.
Semantic tokens have been better explored than acoustic ones in this context.

\subsubsection{Speech Translation}
Among the various SLU tasks, discrete speech tokens are mostly adopted in speech translation, including speech-to-text translation (S2TT) and speech-to-speech translation (S2ST).
Since semantic tokens correlate well with phonetics, they can serve as universal pseudo-labels for untranscribed languages, useful for S2TT in low-resource settings~\cite{zhang-etal-2023-dub}.
Direct S2ST using discrete tokens has garnered more attention on the generation side (Section \ref{sec:application-synthesis}).
Early approaches primarily rely on extracting discrete tokens using VQ-VAEs, particularly for unwritten languages~\cite{tjandra2019speech, zhang2021uwspeech}. 
Recent researches in this area include employing semantic tokens~\cite{lee2022direct, lee-etal-2022-textless, wu2023speechgen}, acoustic tokens~\cite{peng2024mslm, wang-etal-2024-speech, gong2024seamlessexpressivelm}, two-pass architectures~\cite{chen-etal-2023-speech, inaguma-etal-2023-unity}, and non-autoregressive frameworks~\cite{huang2023transpeech}. 
These efforts collectively contribute to advancing the performance and applicability of discrete token-based speech translation systems.

\subsubsection{Unified Speech Understanding}

Discrete tokens provide opportunity to construct unified and adaptable spoken language models with various SLU functionalities.
Efforts include task identifies~\cite{wang2024viola}, prompt tuning~\cite{chang2022speechprompt, chang2023speechprompt, 10.1109/TASLP.2024.3436618,wu2023speechgen}, shared audio and text  vocabulary~\cite{rubenstein2023audiopalm}, and combining continuous and discrete representations~\cite{chen2023lauragpt}.
These efforts highlight the potential of discrete tokens in enhancing the performance and versatility of universal SLU models.

\subsubsection{Limitations}

Despite the advantages and growing popularity in S2ST tasks, discrete tokens still underperform in many SLU tasks. 
Lots of SLU studies~\cite{puvvada2024discrete, chang2024exploring, shon2024discreteslu, cui2024exploring_context, cui2024exploring} only verify that discrete tokens can surpass traditional frequency-domain features in certain tasks such as ASR. 
Continuous SSL features continue to have superior performance~\cite{wang2024comparative}. 
The majority of current LLM-based SLU models rely predominantly on continuous inputs, such as Whisper features~\cite{gong2023joint,chu2023qwen,tang2024salmonn,wavllm,ma2024embarrassingly,bai2024seed}.
Moreover, the performance of discrete tokens in speaker-related tasks is generally much inferior to that of continuous features~\cite{puvvada2024discrete,mousavi2024dasb}.
A significant limitation of discrete tokens for SLU is the inevitable information loss during the quantization process.
Mitigating such loss with more VQ codebooks may hinder the accessibility of semantic information crucial for SLU as well.
Therefore, the full potential of leveraging discrete tokens for SLU remains largely untapped and warrants further exploration.

\subsection{Speech Generation}
\label{sec:application-synthesis}

\subsubsection{Motivation}

Discrete tokens have catalyzed a paradigm shift in speech generation, with TTS being the most representative application.
In TTS systems, discrete tokens are usually used as intermediate features that bridge the acoustic model (text-to-token) and the vocoder or codec decoder (token-to-wav).
There are two major advantages of applying discrete tokens in TTS:
\begin{itemize}
    \item \textbf{Easier training objectives}. Discrete tokens replace the original spectrogram-based regression task with a classification task~\cite{VQTTS}, which can be much easier.
    This also offers a better balance between acoustic models and vocoders, since texts are closer to discrete speech tokens than frequency-domain features.
    \item \textbf{Better use of decoder-only language models}. 
    Decoder-only language models have shown remarkable success in natural language generation.
    After discretization, speech can also be autoregressively generated under the same paradigm. 
    This offers huge potential in leveraging the in-context learning and scaling capabilities of language models to achieve zero-shot high-fidelity TTS~\cite{valle}. 
\end{itemize}
Other generative tasks, such as singing voice synthesis and speech editing, can similarly benefit from the advantages of discrete tokens observed in TTS.
For voice conversion (VC), using discrete tokens as content representations can simplify the process to a token vocoder~\cite{guo2024vec2wav}, when timbre information is effectively removed from the tokens.
Tasks like speech to speech translation~\cite{lee2022direct,zhang2021uwspeech}, speech enhancement~\cite{wang2024selm,liu2024joint} and target speaker extraction~\cite{tang2024tselm} can also be enhanced through language modeling on discrete tokens.

\subsubsection{Autoregressive TTS}
Autoregressively predicting the next VQ index of discrete speech tokens is first proposed in VQTTS~\cite{VQTTS}, which uses an LSTM conditioned on Transformer representations to generate vq-wav2vec~\cite{vq-wav2vec} semantic tokens.
A discrete token vocoder converts the tokens to waveforms with the assistance of handcrafted prosody features.
VQTTS achieves state-of-the-art TTS quality at that time, and shows promising performance in speaker-adaptive TTS~\cite{TNVQTTS,DSETTS,limmits23} and expressive TTS~\cite{liu2024storytts}.

Subsequently, decoder-only TTS models using neural audio codecs have made tremendous success in zero-shot TTS starting from VALL-E~\cite{valle}.
VALL-E contains an autoregressive (AR) model and non-autoregressive (NAR) model, both of which generate EnCodec~\cite{encodec} RVQ tokens.
The AR model performs next-token prediction on the first RVQ layer conditioned on text.
The NAR model predicts the $n+1$-th RVQ tokens given the text, all EnCodec tokens from the speaker reference, and the previous $n$ RVQ layers.
VALL-E employs a concise design in which text and speaker references serve as ``prompts'' for a language model.
It achieves remarkable zero-shot TTS performance when trained on 60k hours of speech.
Later, methods have been proposed to improve generation robustness~\cite{song2024ella,xin2024rall,han2024vall,du2024vall,wang2024attention,chen2024vall}, efficiency~\cite{song24b_interspeech}, style control~\cite{kim2023sc,lyth2024natural,ji2024textrolspeech}, and to incorporate LLMs~\cite{hao2023boosting,shen2024get}.

Besides using an NAR model to predict the rest RVQ layers, alternate modeling strategies have been proposed, such as hierarchical modeling~\cite{yang2024uniaudio} and token interleaving patterns~\cite{musicgen,voicecraft}. 
Semantic tokens are also introduced to cooperate with acoustic codecs~\cite{kharitonov2023speak,vectokspeech,shen2024get,yang2024interleaved}, which might decrease the modeling difficulty since they bridge the gap between texts and acoustics and usually require only a single token stream.
Numerous industry-level large-scale TTS systems have been produced in this autoregressive TTS paradigm, such as XTTS~\cite{casanova2024xtts}, BASE-TTS~\cite{lajszczak2024base}, Seed-TTS~\cite{seedtts}, CosyVoice~\cite{du2024cosyvoice,cosyvoice2}, Fish-Speech~\cite{liao2024fish}, etc.

\subsubsection{Non-Autoregressive TTS}
While autoregressive modeling is the current mainstream of TTS with discrete tokens, non-autoregressive models also exist.
These models either treat the code-vectors as continuous features~\cite{shen2024naturalspeech2}, or directly generate discrete tokens by masked prediction~\cite{wang2024maskgct} or discrete diffusion models~\cite{du2024unicats,facodec}.
These non-autoregressive methods are naturally more robust than autoregressive methods in inference, and also supports speech editing.

\subsubsection{Unified Speech Generation}
The language modeling approach of discrete tokens allows a unified generation framework for multiple tasks.
It suffices to use a task identifier to condition the unified language model.
For example, \cite{vallex,wang2024speechx} extends VALL-E with more tasks like cross-lingual TTS, S2ST, speech editing, etc.
UniAudio~\cite{yang2024uniaudio} supports 11 speech and audio generation tasks within a single hierarchical Transformer model.
Prompt tuning upon a spoken language model has also been explored in \cite{wu2023speechgen} for efficient, transferable and versatile generation.
These efforts demonstrate the potential of a large-scale foundation model for generation.

\subsubsection{Limitations}
In contrast to discrete tokens, another emerging framework for speech generation is diffusion or flow matching-based models, including non-autoregressive models \cite{lee2023hierspeech++,le2024voicebox,liu2024e1,chen2024f5} or autoregressive models \cite{meng2024autoregressive,liu2024autoregressive,zhu2024autoregressive,turetzky2024continuous,jia2025ditar}.
They generate continuous features, and some even eliminate the need for forced alignments in non-autoregressive generation.
Owing to the strong capability of diffusion and flow matching algorithms, they also have remarkable generation fidelity, diversity and controllability.
They can have a higher upper bound for speech quality and intelligibility, as they inherently avoid quantization errors.
In comparison, discrete token-based speech generation models usually fall short in generation robustness.
Therefore, there is an ongoing debate between discrete and continuous representations for speech generation.

\subsection{Text-Free Spoken Language Models}
\subsubsection{Motivation}
End-to-end spoken language and dialogue modeling is one of the most ultimate goals in speech technology.
Discrete tokens are a core component of existing spoken language models, as they enable the language modeling technique to be applied directly on speech. The models discussed in this subsection are text-free spoken language models (TF-SLMs). 
We anticipate that a well-trained TF-SLM will be capable of generating semantically coherent speech without the need for text transcription guidance.

\subsubsection{Existing Efforts}
Ever since GSLM~\cite{lakhotia2021generative} and AudioLM~\cite{borsos2023audiolm} proposed the vision of TF-SLMs, building such models remains a significant challenge even till today. 
This difficulty primarily arises from the lower language modeling efficiency of speech token sequences compared to text, due to their lower semantic information density, longer sequence lengths, and the presence of paralinguistic information~\cite{wang2024whyspeech}. 
Current advancements in TF-SLMs mainly focus on two strategies: (1) reducing token frame rates, and (2) aligning speech with text.

The first approach aims to shorten speech sequences and enhance semantic density by lowering frame rates \cite{lakhotia2021generative, hassid2024textually,shen2024acoustic} to even $\approx$5Hz~\cite{baade2024syllablelm, cho2024sylber}. 
While mitigating sequence length issues to different degrees, they still encounter scalability limitations~\cite{cuervo-marxer-2024-scaling} and compromise reconstruction quality. 
The second strategy involves aligning speech with text through methods like initializing pre-training with text LLMs~\cite{hassid2024textually},  reinforcement learning using ASR and text LLM feedback~\cite{lin2024align}, text-speech token interleaving~\cite{nguyen2024spirit}, adopting novel architectures applied in text language modeling~\cite{park2024longform}, etc~\cite{veluri2024beyond, zhang2024intrinsicvoice}. 
Meanwhile, full duplex modeling has been proposed~\cite{ma2024language} to allow users to interrupt and start new dialogues at will.
However, despite many efforts, these models still struggle to generate semantically reliable long speech during inference due to the lack of explicit transcription guidance.

\subsubsection{Limitations} 
Although these methods show promise, achieving semantic coherence is still a challenging goal, leaving significant progress to be made toward the goal of truly end-to-end spoken language modeling. 
Improving the semantic density and expressiveness of discrete speech representations, making it easier to align text and speech during TF-SLM training, is a promising direction for future exploration.

\subsection{Text-Guided Spoken Language Models}
\subsubsection{Motivation}
Since TF-SLM remains an open problem, the prevalent successful speech dialogue systems settle for an alternative choice that uses text as explicit guidance.
Recent researches, especially following work like OpenAI's GPT-4o\footnote{https://openai.com/index/hello-gpt-4o/}, have focused on SLMs that combine three key capabilities: strong understanding of speech semantics, high-quality speech output, and low latency~\cite{zhang2024speechgptgen, fu2024vita, Xie2024MiniOmniLM, kyutai2024moshi, xie2024mini2, fang2024llamaomni, yu2024salmonnomni, zhong2024lyra, chen2024slamomni, chen2024emova, wang2024freezeomni, zeng2024glm, zhang2024internlm, fu2025vita, luo2025openomni, chen2025minmo}. 
We refer to them as text-guided spoken language models (TG-SLMs). 
Unlike TF-SLMs, while TG-SLMs utilize a unified LLM for seamless processing of user's speech input and system's speech output, they internally decompose the end-to-end speech dialogue process into two well-established sub-procedures: SLU powered by LLMs, and real-time TTS. 
The two sub-procedures are connected via text as an intermediary to stabilize the semantic coherency of the final output. The LLM first generates a textual response to the audio input, then synthesizes the speech token sequence in a streaming fashion. 
In a TG-SLM, the SLU sub-procedure usually uses continuous speech features as input since they preserve more acoustic details for understanding, while the TTS sub-procedure typically uses discrete speech tokens as output to better fit LLM autoregressive generation.

\subsubsection{Speech Generation in TG-SLMs}
To reduce modeling complexity and better align with the autoregressive generation paradigm of LLMs, TG-SLMs favor single-layer discrete speech tokens as direct LLM outputs. 
Existing works make use of either the first layer of an RVQ codec~\cite{zhang2024speechgptgen}, single-codebook codec~\cite{fu2025vita}, or single-codebook supervised semantic token~\cite{zeng2024glm,luo2025openomni}.
Specific designs are introduced corresponding to the tokens, such as chain-of-modality~\cite{zhang2024speechgptgen}, token interleaving to lower latency~\cite{zeng2024glm}, two-stage decoding process~\cite{fu2025vita}, etc.
To better rebuild the speech information with the help of pretrained LLMs, several TG-SLMs use multi-layer speech tokens as LLM output, such as \cite{kyutai2024moshi,Xie2024MiniOmniLM,xie2024mini2}.
They often employ different techniques to generate the text tokens and multi-layer speech tokens in parallel reduce latency.

Mainstream TG-SLMs with discrete tokens as LLM outputs need an additional decoder to synthesize continuous speech signals, either using the codec decoder or a separately-trained vocoder.
There are also efforts to streamingly synthesize speech signals directly based on the LLM hidden embedding~\cite{yu2024salmonnomni,chen2025minmo}, eliminating the need for discrete tokens, additional decoders, or even explicit text tokens, hence further improving the real-time ability.

\subsubsection{Limitations}
Overall, TG-SLMs' task decomposition is effective and flexible.
The SLU sub-procedure can handle both continuous and discrete representations, and single-layer discrete tokens simplify the training and inference of the TTS sub-procedure.
However, unlike TF-SLMs, TG-SLMs rely heavily on text as an intermediary in the TTS sub-procedure, which may overlook paralinguistic information such as emotion, prosody, and environmental context from the previous input, resulting in less coherent and natural response. Additionally, the lack of high-quality annotated conversational data and concerns over security pose significant challenges for the future development of TG-SLMs.

\vspace{-0.05in}
\section{Challenges and Future Directions}
\label{sec:challenge}

Current discrete speech tokens still exhibit certain limitations and challenges that need to be addressed.
In this section, we summarize the existing challenges in this field and outline the corresponding future directions.

\vspace{0.05in}
\subsubsection{Low-Bitrate Tokens}
For bitrates of tokens, factors $Q$ (number of quantizers) and $F$ (frame rate) play a more important role than $V$ (vocabulary size).
Using only a single codebook is very beneficial for language modeling and generation since it frees the need for additional designs for multi-codebook tokens.
A critical problem lies in how to better utilize the highly-compact discrete VQ representation space.
For $F$, the frame rates of most tokens are still much greater than text sequences, which can significantly influence the syntactic and semantic modeling capability of language models~\cite{wang2024whyspeech}.
However, there is usually noticeable performance and intelligibility degradation for tokens with single codebook and small $F$. 
A lower $V$ is also desirable for language modeling and length reduction by BPE.

It remains an open problem what the lower bound of bitrate and the frame rate $F$ are, and how to reach them.
More powerful network architectures or advanced VQ strategies should be helpful, and reducing temporal redundancy by disentangling global information is also a promising solution.

\vspace{0.05in}
\subsubsection{Streaming Ability and Efficiency}
Real-time applications require tokens to be stream-able both in encoding and decoding.
For most CNN-based acoustic tokens, achieving this is easy due to their fixed receptive fields.
For acoustic tokens with Transformer blocks, an attention mask is necessary.
However, most SSL models employ a non-causal Transformer architecture, which makes semantic tokens derived from these models unsuitable for real-time tokenization.
It remains unclear how much performance degradation would result from transitioning to causal architectures in both SSL models and token vocoders.

Streaming ability also poses a requirement for model efficiency.
Currently, larger acoustic token models are reported to achieve better performance with lower bitrates~\cite{xin2024bigcodec,parker2024scalingtransformerslowbitratehighquality}, but at a cost of efficiency.
In addition to reducing the bitrate of the tokenized codes, the efficiency of tokenizers must also be balanced for real-time applications.

\vspace{0.05in}
\subsubsection{Disentanglement in Acoustic Tokens}
Whether disentanglement should be incorporated into acoustic tokens depends on the specific application.
If reconstruction is the major objective, disentanglement may not be necessary.
However, disentanglement can help reduce the bitrate in time-varying tokens, ensure anonymity during transmission, reduce downstream modeling complexity, and achieve independent control of difference voice properties.
There are currently only limited efforts on decoupled acoustic tokens, and the decoupling effect is still suboptimal or causing a negative impact on reconstruction quality.
More advanced techniques for information decoupling should be considered in the future.

\vspace{0.05in}
\subsubsection{Variable Frame Rate Tokens}
As mentioned in Section \ref{sec:variable-rate}, the variable-rate nature of linguistic units can offer an important insight for further reducing the bitrate of tokens, and more importantly, closing the gap between speech tokens and natural language units for downstream tasks.
More explorations need to be taken on variable frame rate acoustic tokens and the benefit of these variable frame rate tokens in practice.

\vspace{0.05in}
\subsubsection{Combining Acoustic and Semantic Tokens}
Given the distinct properties of acoustic and semantic tokens, a natural question arises: 
Can a representation space contain rich speech understanding capabilities while also reconstructing acoustic details at a decent level?
Incorporating semantic information from SSL models has proven to enhance the reconstruction and downstream modeling performance of acoustic tokens~\cite{zhang2024speechtokenizer,ye2024codec,liu2024semanticodec}.
Recently, explicit text supervision has also sparked remarkable progress in acoustic tokens~\cite{qiang2025secousticodec,wang2025tadicodec,gong2025xy}.
We anticipate more promising results in this direction.

\vspace{0.05in}
\subsubsection{Paralinguistics in Semantic Tokens}
While speaker information is generally considered irrelevant for semantic content, prosody serves as a crucial component of paralinguistic information.
Semantic tokens derived through simple clustering methods are likely to discard both speaker information and prosody, harming downstream models' ability to handle rich emotions, tones, singing voices, and non-verbal vocalizations that convey semantic meaning.
This problem can be partially mitigated by certain VQ approaches that encode more information from SSL features~\cite{huang2023repcodec,shi24h_interspeech,mousavi2024should}, but at a cost of more codebooks and higher bitrates.
Supervised tokenization could also be considered for directly guiding tokens toward paralinguistic information in the future. 

\vspace{0.05in}
\subsubsection{Noise Preservation vs. Noise Robustness}
Similar to disentanglement in acoustic tokens, the inclusion or exclusion of background noise and channel effects in the tokens also depends on the specific application.
Most acoustic tokens are designed to capture noise, but their performance across various types of noise and channel effects remains unclear.
This issue extends beyond speech and relates to the broader scope of neural \textit{audio} codecs.
On the other hand, denoising~\cite{zeghidour2021soundstream} is also an interesting application of acoustic tokens that leverages the limited VQ space.
If noise is considered undesirable in tokens, such as semantic tokens, then the robustness against various types of signal perturbations needs to be investigated.

\vspace{0.05in}
\subsubsection{Timbre Control in Token Vocoders}
For semantic tokens and speaker-decoupled acoustic tokens, token vocoders should be responsible for controlling speaker timbre.
Currently, both GAN-based token-to-wav vocoders~\cite{guo2024vec2wav} and flow matching-based token-to-mel models~\cite{du2024cosyvoice} have demonstrated strong timbre control capabilities. 
It remains an open question whether the upper bound of the former method can be improved by training on large-scale datasets, as is done with the latter.
Also, the timbre controllability of in-the-wild reference prompts with various acoustic conditions should be further investigated.



\vspace{-0.07in}
\section{Conclusion}
Recently, discrete speech tokens have emerged as a rapidly evolving field and a core research direction in the speech LLM era. These tokens encode acoustic or semantic information into a compact discrete representation space, catalyzing the fusion of LLMs and speech processing.
In this review, we provide a comprehensive introduction to representative categories of discrete speech tokens, summarizing their motivations and limitations. 
We conduct a unified analysis of reconstruction, voice conversion, and downstream semantic modeling across different token types to highlight their unique characteristics. 
We also review popular applications of discrete tokens in speech processing tasks, including understanding, generation and language modeling of speech. 
Finally, we explore future directions for discrete speech tokenization methods. 
We hope this review lays a solid foundation for future research in speech technology.

\vspace{-0.1in}
\section*{Acknowledgments}
We thank Haoran Wang, Jingyu Zhou, and Shuai Wang for their contribution in a tutorial related to this review paper.
\vspace{-0.1in}

\bibliographystyle{IEEEtran}
\bibliography{refs}

\clearpage
\setcounter{section}{0}  
\renewcommand{\thesection}{Appendix~\Roman{section}}  

\section{List of Widely-Used Discrete Speech Tokens}
To give a high-level overview of existing discrete speech tokens, we summarize the widely-used discrete acoustic tokens in Table \ref{tab:acoustic-metadata}, and semantic tokens in Table \ref{tab:semantic-metadata}.

\begin{table*}[]
\centering
\caption{A summary of widely-used acoustic speech tokens (neural speech codecs). 
Note that some codecs in this list are also applicable to general audio.
Italic ``\textit{C,T,U}'' denote CNN, Transformer or U-Net-based generator architecture.
Symbols `/' and `-' denote ``or'' and ``to'' for different model versions, ``+'' means different configurations in different VQ streams in a single model, and ``$\approx$'' means the average value.
$Q,F,V$ mean number of quantizers, frame rate and vocabulary size of each quantizer respectively. 
For example, ``$Q=2$, $V$=8192+(4096-32768)'' in SemantiCodec means one of the two VQ streams has 8192 possible codes, and the other can vary from 4096 to 32768 in different configurations.
For FSQ, $V$ is presented as $L^d$ where $L$ is the quantization levels for each dimension, and $d$ is the number of dimensions.
When $L$ is different for each dimension, like LFSC, we use $L_1\times L_2\times \cdots$ to represent each dimension.
``Dynamic'' means the number for each token frame is variable, like in variable-bitrate and variable frame rate setups.
Bitrates are computed by $\frac1{1000}\sum_{i=1}^Q F_i\lceil \log_2 V_i\rceil$ kbps, without entropy coding. }
\label{tab:acoustic-metadata}
\resizebox{\textwidth}{!}{
\begin{tabular}{@{}lccccccc@{}}
\toprule
\textbf{{Acoustic Speech Tokens}} & \textbf{Model Framework} & \textbf{\makecell{Sampling\\Rate (kHz)}} & \textbf{\makecell{Quantization\\Method}} & \textbf{$Q$} & \textbf{$F$ (Hz)} & \textbf{$V$} & \textbf{{Bitrate (kbps)}}  \\ \midrule
\multicolumn{6}{l}{\textbf{\textit{General-purpose acoustic tokens}}} \\
SoundStream~\cite{zeghidour2021soundstream} & VQ-GAN (\textit{C}) & 24 & RVQ & max 24 & 75 & 1024 & max 18.00 \\
EnCodec~\cite{encodec} & VQ-GAN (\textit{C})& 24 & RVQ & max 32 & 75 & 1024 & max 24.00  \\
TF-Codec~\cite{jiang2023latent} & VQ-GAN (\textit{C}) & 16 & GVQ & 3-32 & 25 & 512 / 1024 & 0.68-8.00 \\
Disen-TF-Codec~\cite{jiang2023disentangled} & VQ-GAN (\textit{C}) & 16 & GVQ & 2 / 6 & 25 & 256 / 1024 & {0.40 / 1.50} \\
AudioDec~\cite{audiodec} & VQ-VAE (\textit{C})+GAN& 48 & RVQ & 8 & 160 & 1024 & 12.80 \\
HiFi-Codec\cite{yang2023hifi} & VQ-GAN (\textit{C})& 16 / 24 & GRVQ & 4 & 50-100 & 1024 & 2.00-4.00 \\
DAC~\cite{kumar2024high} & VQ-GAN (\textit{C})& 44.1 & RVQ & 9 & 86 & 1024 & 7.74 \\
LaDiffCodec~\cite{yang2024generative} & Diffusion & 16 & RVQ & 3 / 6 & 50 & 1024 & 1.50 / 3.00 \\
{FreqCodec}~\cite{du2024funcodec} & VQ-GAN (\textit{C})& 16 & RVQ & max 32 & 50 & 1024 & max 16.00 \\
TiCodec~\cite{ticodec} & VQ-GAN (\textit{C})& 24 & RVQ, GVQ & 1-4 & 75 & 1024 & 0.75-3.00  \\
APCodec~\cite{APCodec} & VQ-GAN (\textit{C})& 48 & RVQ & 4 & 150 &1024 & 6.00 \\
SRCodec~\cite{zheng2024srcodec} & VQ-GAN (\textit{C}) & 16 & GRVQ   & 2-8 & 50 & 512+1024 & 0.95-3.80\\
SQ-Codec~\cite{yang24l_interspeech} & VQ-GAN (\textit{C})& 16 & FSQ & 1 & 50 &$19^{32}$ & 8.00 \\
Single-Codec~\cite{singlecodec} & VQ-GAN (\textit{T+C})& 24 & VQ & 1 & 23 & 8192 & 0.30  \\
ESC~\cite{gu2024esc}  & VQ-GAN (\textit{U})& 16 & GVQ & max 18 & 50 & 1024 & max 9.00 \\
CoFi-Codec~\cite{guo2024speaking} & VQ-GAN (\textit{U}) & 16 & GVQ & 3 & 8.33+25+50 & 16384 & 1.17 \\
HILCodec~\cite{ahn2024hilcodec} & VQ-GAN (\textit{C})& 24 & RVQ & 2-12 & 75 & 1024 & 1.50-9.00 \\
SuperCodec~\cite{zheng2024supercodec} & VQ-GAN (\textit{C})& 16 & RVQ & 2-12 & 50 & 1024 & 1.00-6.00\\
SNAC~\cite{Siuzdak_SNAC_Multi-Scale_Neural_2024} & VQ-GAN (\textit{C})& 24 & RVQ & 3 &12+23+47 & 4096 & 0.98 \\
WavTokenizer~\cite{ji2024wavtokenizer} & VQ-GAN (\textit{C})& 24 & VQ & 1 & 40 / 75 & 4096 & 0.48 / 0.90  \\
BigCodec~\cite{xin2024bigcodec} & VQ-GAN (\textit{C})& 16 & VQ & 1 & 80 & 8192 & 1.04 \\
LFSC~\cite{casanova2024low} & VQ-GAN (\textit{C})& 22.05 &  FSQ & 8 & 21.5 & $8\times7\times6\times6$ & 1.89 \\
NDVQ~\cite{niu2024ndvq} & VQ-GAN (\textit{C})& 24 &  RVQ & max 32 & 75 & 1024 & max 24.00 \\
VRVQ~\cite{chae2024variable} & VQ-GAN (\textit{C})& 44.1 & RVQ & dynamic, max. 8 & 86 & 1024 & {0.26 + max 6.89}\\
TS3-Codec~\cite{wu2024ts3codectransformerbasedsimplestreaming} & VQ-GAN (\textit{T})& 16 & VQ & 1 & 40 / 50 & 65536 / 131072 & 0.64-0.85 \\
Stable-Codec~\cite{parker2024scalingtransformerslowbitratehighquality} & VQ-GAN (\textit{T})& 16 & FSQ  & 1 / 2 & 25 & $5^6$ / $6^6$ & 0.40 / 0.70 \\
FreeCodec~\cite{zheng2024freecodecdisentangledneuralspeech} & VQ-GAN (\textit{C+T})& 16 & VQ & 1+1 & 50+7 & 256 & 0.45  \\
{FocalCodec}~\cite{della2025focalcodec} & VQ-GAN (\textit{T+C}) & 16 & FSQ & 1 & 12.5 / 25 / 50 & $2^{13}$  & 0.16 / 0.33 / 0.65\\
ALMTokenizer~\cite{yang2025almtokenizer} & VQ-GAN (\textit{T}) & 24 & RVQ & 3 & 12.5 & 2048 & 0.41 \\
{TFC}~\cite{zhang25k_interspeech_tfc}  & VQ-GAN (\textit{C)} & 24 & RVQ & 8 & dynamic, 18.75-75 & 1024 & 1.50-6.00\\
{CodecSlime}~\cite{wang2025codecslime} & VQ-GAN (\textit{C}) & 16 & FSQ & 1 & dynamic, $\approx$40 & $5^2\times 3^6$ & 0.08 + ($\approx$0.60) \\
\midrule
\multicolumn{6}{l}{\textbf{\textit{Mixed-objective acoustic tokens: semantic distillation}}} \\
{Siahkoohi et al.}~\cite{siahkoohi22_interspeech} & VQ-GAN (\textit{C})& 16 & RVQ & 2+1 / 2+2 / 6 & 25+50 & 64 & 0.60 / 0.90 / 1.80\\
SpeechTokenizer~\cite{zhang2024speechtokenizer} & VQ-GAN (\textit{C})& 16 & RVQ & 8 & 50 & 1024 & 4.00 \\
SemantiCodec~\cite{liu2024semanticodec} & Diffusion & 16 & VQ & 2 & 12.5-50 & {8192+(4096-32768)} & 0.31-1.40 \\
LLM-Codec~\cite{yang2024uniaudio15} & VQ-GAN (\textit{C})& 16 & RVQ & 3 & 8.33+16.67+33.33 & 3248+32000+32000& 0.85 \\
X-Codec~\cite{ye2024codec} & VQ-GAN (\textit{C})& 16 & RVQ & max 8 & 50 & 1024 & max 4.00 \\
SoCodec~\cite{guo2024socodec} & VQ-GAN (\textit{C})& 16 & GVQ & 1 / 4 / 8 & 25 / 8.3 / 4.2 & 16384 & 0.35 / 0.47 \\
Mimi~\cite{kyutai2024moshi} & VQ-GAN (\textit{C+T})& 24 & RVQ & 8 & 12.5 & 2048 & 1.10 \\
X-Codec 2.0~\cite{ye2025llasa} & VQ-GAN (\textit{C+T}) & 16 & FSQ & 1 & 50 & $4^8$ & 0.80 \\
{BiCodec}~\cite{wang2025sparktts} & VQ-GAN (\textit{C}) & 16 & VQ & 1 & 50 & 8192 & 0.65 \\ 
{DualCodec}~\cite{li25e_dualcodec} & VQ-GAN (\textit{C}) & 24 & RVQ & 3 / 6 & 12.5 / 25 & (1024 / 16384)+(1024 / 4096) & 0.75-0.93 \\
XY-Tokenizer~\cite{gong2025xy} & VQ-GAN (\textit{T)} & 16 & RVQ & 8 & 12.5 & 1024 & 1.00 \\
TaDiCodec~\cite{wang2025tadicodec} & Diffusion & 24 & FSQ & 1 & 6.25 & $2^{14}$ & 0.0875 \\ 
\midrule
\multicolumn{6}{l}{\textbf{\textit{Mixed-objective acoustic tokens: disentanglement}}} \\
SSVC~\cite{SSVC} & VQ-GAN (\textit{C})& 24 & RVQ & 4 & 50 & 512 & 1.80 \\
PromptCodec~\cite{pan2024promptcodec} & VQ-GAN (\textit{C})& 24 & GRVQ & 1-4 & 75 & 1024 & 0.75-3.00 \\
FACodec~\cite{facodec} & VQ-GAN (\textit{C})& 16 & RVQ & 1+2+3 & 80 & 1024 & 4.80 \\
LSCodec~\cite{guo2024lscodec} & VQ-VAE (\textit{C+T})+GAN& 24 &  VQ & 1 & 25 / 50 & 1024 / 300& 0.25 / 0.45 \\
SD-Codec~\cite{bie2024learning} & VQ-GAN (\textit{C})& 16 &  RVQ & 12 & 50 & 1024 & 6.00 \\
{DeCodec}~\cite{luo2025decodec}  & VQ-GAN (\textit{C)} & 16 & RVQ & 8+8 & 50 & 1024 & 8.00\\ 
\bottomrule
\end{tabular}
}
\end{table*}

\begin{table*}[]
\centering
\caption{A high-level summary of widely-used semantic speech tokens. Notations follow Table.\ref{tab:acoustic-metadata}.
Symbol `/' denotes different versions. 
``Inner Quantizer'' refers to whether the model has a quantizer, or external quantization (e.g. clustering) must be performed.
$F$ denotes frame rate.
In case there are inner quantizers, $Q,V$ denote number of quantizers and vocabulary size for each quantizer, respectively.
``\textit{NR}.'' means not reported.
}
\label{tab:semantic-metadata}
\begin{tabular}{@{}lcccccc@{}}
\toprule
\textbf{\makecell{Semantic Speech Tokens}} & \textbf{\makecell{Criterion / Objective}} & \textbf{\makecell{Training Data (h)}} & $F$ \textbf{(Hz)} & \textbf{{Inner Quantizer}} \\ \midrule
\multicolumn{5}{l}{\textbf{\textit{From self-supervised learning (SSL) models}}} \\
vq-wav2vec~\cite{vq-wav2vec} & Contrastive & 0.96k & 100 & GVQ, $Q=2,V=320$ \\
wav2vec 2.0~\cite{baevski2020wav2vec} & Contrastive & 60k & 50 & GVQ, $Q=2,V=320$ \\
XLSR-53~\cite{conneau21_interspeech} & Contrastive & 50k & 50 & GVQ, $Q=2,V=320$ \\
HuBERT~\cite{hsu2021hubert} & Predictive & 60k & 50 & No \\
WavLM~\cite{chen2022wavlm} & Predictive & 94k & 50 & No \\
BEST-RQ~\cite{chiu2022self} & Predictive & 60k & 25 & {No} \\ 
w2v-BERT~\cite{chung2021w2v} & {Predictive+Contrastive} & 60k & 50 & VQ, $Q=1,V=1024$ \\
w2v-BERT 2.0~\cite{barrault2023seamless} & {Predictive+Contrastive} & 4500k & 50 & GVQ, $Q=2,V=320$ \\
DinoSR~\cite{liu2024dinosr} & Predictive & 0.96k & 50 & VQ, $Q=8,V=256$ \\
NEST-RQ~\cite{han2024nest} & {Predictive} & 300k &  25 & {No} \\
LAST~\cite{turetzky2024last} & {Predictive} & \textit{NR.} & 50 & VQ, $Q=1,V=500$ \\
{Gat et al.~\cite{gat2023augmentation}} & Noise Invariance & 0.10k & 50 & VQ, $G=1,V=50$-$500$  \\
ContentVec~\cite{qian2022contentvec} & Speaker Invariance & 0.96k & 50 & No \\
SPIRAL~\cite{huang2022spiral} & Noise Invariance & 60k & 12.5Hz & No\\
CCC-wav2vec 2.0~\cite{ccc-wav2vec2.0} & Noise Invariance & 0.36k & 50 & GVQ, $G=2,V=320$ \\
Spin~\cite{chang23_interspeech} & Speaker Invariance & 0.10k & 50 & VQ, $Q=1,V=128$-$2048$\\
NAST~\cite{messica2024nast} & Noise Invariance & 0.96k & 50 & VQ, $Q=1,V=50$-$200$\\
DC-Spin~\cite{chang2024dc} & Speaker Invariance & 0.96k & 50 & VQ, $Q=2,V=(50$-$500)$+$4096$ \\
\midrule
\multicolumn{5}{l}{\textbf{\textit{Supervised semantic tokens}}} \\
$\mathcal S^3$ Tokenizer~\cite{du2024cosyvoice}  & Supervised ASR & 172k & 25 / 50  & VQ, $Q=1,V=4096$ \\
Zeng et al.~\cite{zeng2024scaling} & Supervised ASR & 90k & 12.5 & VQ, $Q=1,Q=16384$ \\
Du et al. (CosyVoice 2)~\cite{cosyvoice2} & Supervised ASR & 200k & 12.5 & FSQ, $Q=1,V=3^8$ \\
{Du et al. (CosyVoice 3)\cite{du2025cosyvoice3}} & Supervised Multi-Task &  530k & 25 & FSQ, {Details \textit{NR.}}\\
\bottomrule
\end{tabular}
\vspace{-0.15in}
\end{table*}

\section{Experimental Details}


\subsection{Vocoder for Reconstruction and Voice Conversion}
In the experiments of reconstruction and voice conversion, we train a speech token vocoder for each of the semantic tokens.
This speech token vocoder is chosen as CTX-vec2wav$^\alpha$~\cite{guo2024lscodec} as the improved version of CTX-vec2wav~\cite{du2024unicats}.
This vocoder is a timbre-controllable speech token vocoder.
The input speech tokens are converted to their corresponding code-vectors first, where code-vectors are concatenated along channel dimension when there are multiple codebooks.
Then, these code-vectors are passed to a Conformer~\cite{conformer} frontend before a HifiGAN~\cite{kong2020hifigan} generator module.
The timbre information is provided to the vocoder model from a reference waveform.
This reference is passed to a WavLM-Large~\cite{chen2022wavlm} model to extract strong timbre-dependent embeddings.
These embeddings are extracted from the 6-th layer of this WavLM model, following previous work~\cite{knnvc}.
Then, they are fed to the Conformer frontend via position-agnostic cross-attention mechanism~\cite{du2024unicats}.

The training of these CTX-vec2wav$^\alpha$ vocoders follows the GAN-based vocoder paradigm where a set of discriminators are employed.
We follow the detailed training setup in this repository\footnote{\url{https://github.com/cantabile-kwok/vec2wav2.0}}.
The generator of this vocoder consists of $\approx$32M parameters.
We train the vocoder for each token up to 1 million steps on 4 GPUs, with a dynamic batch size of $\approx$36s of speech per GPU.
The training data is all LibriTTS training splits, which amount to about 585 hours.

After obtaining a vocoder for each semantic token, we can perform reconstruction comparisons on all tokens, and voice conversion comparisons on semantic tokens and some acoustic tokens.
For reconstruction, we provide the original utterance as the source of timbre information. 
One may argue that this setup carries a risk of information leakage, since content information may also be implicitly encoded in timbre representations. 
An alternative strategy is to use another reference prompt from the same speaker to supply timbre information. 
However, this essentially becomes a ``same-speaker conversion'' task rather than reconstruction, and the resulting metrics are not strictly comparable to those of ordinary acoustic tokens that do not require additional speaker inputs.
Given that the CTX-vec2wav$^\alpha$ vocoder has never been trained on cases where timbre representations and content tokens originate from the same segment, we consider this strategy to still provide sufficiently fair comparisons.

The details in evaluation metrics include:
\begin{itemize}
    \item For WER, we transcribe the synthesized speech using NeMo-ASR\footnote{\url{https://huggingface.co/nvidia/stt_en_fastconformer_transducer_large}}, normalize the text using Whisper normalizer\footnote{\url{https://github.com/openai/whisper/tree/main/whisper/normalizers}}, and use \texttt{jiwer}\footnote{\url{https://github.com/jitsi/jiwer}} to measure the total WER.
    \item For GPE and P.Corr, we use the YIN algorithm in PyWorld\footnote{\url{https://github.com/JeremyCCHsu/Python-Wrapper-for-World-Vocoder}} to extract pitch contours. Pitch errors and correlation are only computed on frames where both the reference and the synthesized waveforms are voiced.
    \item For PESQ, we compute wide-band PESQ using  this tool\footnote{\url{https://github.com/ludlows/PESQ}}. For STOI, we use this tool\footnote{\url{https://github.com/mpariente/pystoi}}.
    \item For SECS, we use Resemblyzer\footnote{\url{https://github.com/resemble-ai/Resemblyzer}} to extract embeddings for both the reference waveform and the synthesized waveform. Then, SECS is computed as the cosine similarity between the two embeddings.
\end{itemize}

The test metadata for reconstruction and voice conversion can be found here\footnote{\url{https://cpdu.github.io/unicats/resources/testsetB_utt2prompt}} and here\footnote{\url{https://cantabile-kwok.github.io/LSCodec/audio_ready/VC/utt2prompt}}, respectively.

\subsection{Downstream Semantic Modeling}

\subsubsection{Motivation}

Given that a large branch of discrete speech tokens is semantic tokens, it is also important to measure the semantic modeling abilities of speech tokens in downstream tasks, without relying on a codec decoder or vocoder to reconstruct into waveforms.
There are mainly two approaches to probe the semantic modeling abilities in discrete speech tokens:
\begin{itemize}
    \item \textit{Index-based}: Like DASB~\cite{mousavi2024dasb}, discrete indices are directly used as inputs and passed to a set of learnable embedding lookup modules before subsequent networks.
    When there are multiple code streams (for example, when using GVQ or RVQ), a learnable attention weight will aggregate the embeddings from different code streams.
    \item \textit{Vector-based}: The code-vectors corresponding to the discrete indices are considered as inputs to the probing network.
    This is a direct extension to the SUPERB~\cite{superb} benchmark that mainly considers continuous speech representations.
\end{itemize}
The \textit{vector-based} approach, in other words, probes the information from the established latent space where speech tokenizers perform quantization on.
Meanwhile, the \textit{index-based} approach concentrates on the abstract discrete indices and anticipates that indices themselves carry important semantic information rather than the underlying high-dimensional space.
Since the most of the discrete token-based speech generation models use token indices instead of code-vectors as modeling targets, we opt to use the \textit{index-based} approach in this survey.

To better align with the results in previous works, we use the DASB~\cite{mousavi2024dasb} framework and its official implementation\footnote{\url{https://github.com/speechbrain/benchmarks/tree/DASB/benchmarks/DASB}}.

\subsubsection{Tasks and Datasets}
We consider two of the most representative speech semantic tasks: automatic speech recognition (ASR) and intent classification (IC).
In ASR, we use the characters as text units for simplicity, and the connectionist temporal classification (CTC)~\cite{ctc} criterion for training.
We use all the 960 hours in LibriSpeech~\cite{librispeech} as the training data, dev-clean split as the validation set, and test-clean as the test set.
IC is the process of understanding the underlying goal or purpose of a user's spoken utterance.
We use SLURP~\cite{bastianelli-etal-2020-slurp} for the IC task, which includes 18 classification categories, such as event name, date, etc.
We use the train-real split for training and the official dev and test sets, which contains about 40, 7 and 10 hours of speech data, respectively.

All the original waveforms are in 16kHz. If a speech tokenizer requires input at a different sampling rate, we resample the waveform before feeding them accordingly.

\subsubsection{Model}
Following DASB, we use LSTM-based networks for both tasks.
For each token, the input layer consists of $Q$ embedding lookup tables of shape $V \times 1024$, where $Q$ is the number of codebooks and $V$ is the vocabulary size of each codebook. 
For each frame, the $Q$ embeddings are fed into an multi-layer perceptron (MLP) with a hidden layer of  1024 dimensions, to produce softmax probabilities that weight-sum the $Q$ embeddings.
Then, an bidirectional LSTM with 2 layers and 1024 hidden neurons takes the embedding sequence as input, and produces 2048-dimensional outputs.
For the ASR task, another linear layer is applied to produce emission probabilities on the 31 text units (26 English characters and special tokens).
For the IC task, mean pooling over the sequence axis is applied before a linear projection to the logits on 18 output classes.
Besides the input embedding layer which is dependent on the specific token configuration, the model has around 43M parameters.

It is important to note that some speech tokens have a low frame rate, such as 12.5Hz for Mimi~\cite{kyutai2024moshi}. As the character rate is about 14Hz, the original low frame rates make these tokens not suitable for character-based CTC ASR.
Hence, when the token frame rate is less than 50Hz, we always repeat the input tokens to at least 50Hz for ASR task.
For the IC task, no repeating is performed.

\subsubsection{Training and Decoding}
For the ASR task, we use AdamW~\cite{loshchilov2018decoupled} optimizer with initial learning rate $10^{-4}$ and weight decay $5\times 10^{-4}$. 
For the IC task, we use Adam~\cite{adam} optimizer with initial learning rate $2\times 10^{-4}$.
For both tasks, the learning rate is reduced to 80\% of its original value whenever the improvement in validation loss after an epoch is less than 0.25\% relative to the previous epoch. For each token, training is conducted for 20 epochs on a single GPU, and the checkpoint with the lowest validation loss is selected for testing.

In ASR task, we use beam search decoding with beam size 100.
Other hyperparameters all remain consistent with the DASB official configuration.
For some tokens, training suffers from collapse, leading to  repetitive and meaningless characters in decoding.
This is usually observed when the token has a large vocabulary size (like 2 groups of 15625 FSQ indices for StableCodec~\cite{parker2024scalingtransformerslowbitratehighquality}, and 2 groups of 32000 VQ indices for LLM-Codec~\cite{yang2024uniaudio15}).
Following DASB, we denote these cases as ``not converged'' (NC), since the metrics are not meaningful there.

\newpage

\vfill

\end{document}